**On Casimir force between two real metallic plates**

M.V. Davidovich

*Saratov State University,*

*410012 , Saratov, Russia*

E-mail: davidovichmv@yandex.ru

The corrections to Casimir pressure (force) on two plates of real metal described by the dielectric constant of the Drude, Drude-Lorentz, Drude-Smith and plasma models at the finite temperature and conductivity are considered. We consider the contour integral lost in the Lifshitz formula and discuss inconsistencies for the corrections to the Casimir force given in the literature. Numerical and analytical results in the near and far zones are presented. The two first-order correction obtained due to the actual dielectric constant of the metal coincided with the literature data.

## 1. Introduction

The imposition or change of boundary conditions by bodies located in a vacuum leads to a change in the energy $\hbar\omega/2$ of zero fluctuations in the quantum vacuum, which leads to the appearance of dispersion forces, which are a type of van der Waals forces and act on bodies at distances of the order of 1 nm or more. These are macroscopic forces caused by the dielectric permittivity (DP) of bodies. This was first shown by Casimir in 1948, by examining the interaction between two ideally conducting mirrors with a gap *d*. For them, the force density varies with distance according to the law $1/d^4$. With an infinitesimal gap, the energy density in it is zero, and outside the gap one is infinite, which leads to an infinite increase in force. For real metals, the surface conductivity $\sigma(\omega)=i\omega\varepsilon_0(\varepsilon(\omega)-1)$ tends to infinity as the frequency $\omega$ tends to zero, since the negative imaginary DP of the metal $\varepsilon(\omega)=-i\omega_p^2/(\omega\omega_c)\to\infty$ tends to infinity (for the dependence $\exp(i\omega t)$, $\varepsilon(\omega)=\varepsilon'(\omega)-i\varepsilon''(\omega)$, $\varepsilon''(\omega)>0$ ).

This means that for metals, Casimir's law holds in the far zone $d>>\lambda_p=2\pi c/\omega_p$, and, generally speaking, in the limit $d\to\infty$, which requires determining corrections for a finite *d*. For metals, the plasma wavelength is $\lambda_p$ ~ 120 nm, and in the far zone of $\lambda_p << d << \lambda_c$, the force is determined by the plasma model (PM) for DP: $\varepsilon=1-\omega_p^2/\omega^2$. When determining the force in a zone of even greater distances, the PM stops working, and the Drude model (DM) should be used for DP: $\varepsilon = \varepsilon_L - \omega_p^2 / \left(\omega^2 - i\omega\omega_c\right)$. The collision frequency (CF) $\omega_c$ and plasma frequency (PF) $\omega_p$ are indicated here. At low frequencies $\omega << \omega_c$ for DM $\varepsilon(\omega) = -i\omega_p^2 / (\omega\omega_c) = -i\sigma(0)\varepsilon_0 / \omega$ . For pure metals, when the temperature tends to zero, the CF tends to zero as $\omega_c = \omega_{c0} T / T_0$ ($T_0$ is the room temperature and $\omega_{c0}$ is the CF for it), therefore, MD turns into PM. The same applies to extremely high frequencies. In the near range, the DP of metals is determined by its properties in the UV range, i.e. by the Lorentz dispersion with many terms. Therefore, the $1/d^4$ law stops working. In the hard UV and X-ray ranges, the metal becomes transparent, its DP $\varepsilon \to 1$, and the force density should be determined by the van der Waals method, for example, using the density functional theory and the interaction potentials of matter atoms. In the range of distances from 1 to 120 nm, the Lorenz (Drude-Lorentz) model should be used. When determining the force numerically, experimental data for the imaginary part of DP can be used and recalculated using the Kramers-Kronig ratios.

In addition to zero vacuum fluctuations, the existing fluctuations in the thermal field lead to fluctuations in the polarization currents of bodies and to the appearance of dispersion forces related to temperature. Their contribution is usually significantly less than the contribution of zero fluctuations. The forces are essentially determined by the spectral properties of the materials of the bodies: their DP and conductivities. Next, for DP, we consider PM, DM, the Drude-Smith model (DSM), as well as the internal field models: Lorenz-Lorentz and Onsager. For these models, we compare the dispersion forces.

In [1–3] Lifshitz gave a formula for the external Casimir pressure $P_C$ (the force of attraction per unit area) $P(d) = -F(d)/S = -\tilde{F}$ for two thick (semi-infinite) dielectric plates with DP ε separated by a gap $d$, in the form (see also [4]):

$$\tilde{F}(d) = \frac{\hbar}{2\pi^2 c^3} \int_0^\infty \xi^3 d\xi \int_1^\infty \sum_{\mu=e,h} \frac{r_\mu^2 \exp(-2p\xi d/c)}{1 - r_\mu^2 \exp(-2p\xi d/c)} p^2 dp \,. \tag{1}$$

$\tilde{F}$ is specific gravity. External pressure means attraction (negative internal pressure in the gap). The external pressure means that the energy density outside the plates is greater than in the gap between them. In this formula, the plates are assumed to be the same, the temperature is zero, and the squares of the reflection coefficients (RC) $r_\alpha^2$ of E–waves ($p$-polarization) and H-waves ($s$-polarization) from the plates on the gap side are: $r_e = (s - \varepsilon p)/(s + \varepsilon p)$, $r_h = (s - p)/(s + p)$, $s = \sqrt{p^2 - 1 + \varepsilon}$, and $\xi = i\omega$ means the imaginary frequency. Note that dependence $\exp(i\omega t)$ is used in this paper as usual in radio engineering. To avoid misunderstandings during the transition to dependence $\exp(-i\omega t)$ (as in [1–3]), the replacement $i \to -i$ should be made. Next, we also use an imaginary wave number $k = ik_0 = i\omega / c$. In [2,3], a formula for the final temperature was proposed

$$P(d,T) = -\frac{k_B T}{\pi}\sum_{n=0}^{\infty}\frac{k_n^3}{1+\delta_{n0}}\int_1^{\infty}\sum_{\mu=e,h}\frac{p^2 dp}{r_{\mu n}^{-2}\exp(2pk_n d)-1}, \tag{2}$$

In which $k_n = 2\pi k_B T n/(\hbar c) = 2\pi n/\delta_T$, $r_{\mu m} = r_\mu(k_n)$. In the monograph [4], in addition to these two formulas, the following form of formula (2) is also given

$$\tilde{F}(d,T) = \frac{\hbar}{2\pi^2}\int_0^{\infty} d\omega \coth(x) \mathrm{Re}\int_0^{\infty}\kappa k_z \sum_{\mu=e,h}\frac{r_\mu^2 \exp(-2ik_z d)}{1-r_\mu^2\exp(-2ik_z d)}d\kappa, \tag{3}$$

in which $k_z = \sqrt{k_0^2-\kappa^2} = -i\tilde{k}_z$, $\tilde{k}_z = \sqrt{\kappa^2 - k_0^2}$. $P = -\tilde{F}$ is the external Casimir pressure, the integration takes place over the frequency ω and the variable $\kappa = \sqrt{k_x^2 + k_y^2}$, $x = \hbar\omega/(2k_B T) = \hbar k_0 c/(2k_B T)$, $k_0 = \omega / c$. However, this integral diverges and does not lead to (2). In [3], the next formula (4) is given as the initial one, obtained as if using the Rytov method of introducing fluctuation sources into Maxwell's equations and by decomposing into plane waves $A(\omega,\mathbf{k})\exp(-ik_x x - ik_y y - ik_z z)$ with the imposition of boundary conditions:

$$P(d,T) = -\frac{\hbar}{2\pi^2 c^3}\int_0^{\infty}\omega^3 d\omega \coth(x) \mathrm{Re}\int\sum_{\mu=e,h}\frac{p^2}{r_\mu^{-2}(\omega,p)\exp(2ip\omega d/c)-1}dp. \tag{4}$$

Formula (3) from [4] seems to follow from (4), but this is not the case. It is assumed that the integral function in (4) is analytical except for the poles (although the real part is taken from it) and then it is transformed to (2) by changing the integration contour without crossing the poles. In this case, the value of *p* is considered to be complex, varying from one to zero along the real axis when changing $0 < \kappa < k_0$, and from 0 to $i\infty$ on the imaginary axis when $k_0 < \kappa < \infty$. The first case corresponds to radiated modes, and the second case corresponds to surface modes. Next, (4) is transformed by rotating the Wick (see also [4]) to the integral along the imaginary frequency axis and along the real axis for the variable *p*. Let us do this for $r_\mu^{-2} = 1/$ According to [3], the inner integral in the form\

$$\frac{1}{2}\int_0^1 p^2 dp(1-\exp(-2ip\omega d/c)) - i\int_0^{i\infty}\frac{p^2 dp}{\exp(-2p\omega d/c)-1}.$$

The first integral contains a frequency-independent term 1/6, so the frequency integral diverges. The second integral can be represented as

$$\mathrm{Re}\sum_{m=1}^{\infty}\exp(2mp\omega d/c)\left[\frac{ip^2}{(2m\omega d/c)} - \frac{2ip}{(2m\omega d/c)^2} + \frac{i}{(2m\omega d/c)^3}\right]_0^{i\infty}.$$

It should give zero if we assume that the rapidly oscillating exponent is zeroing out. It must match the surface modes, which are not available for ideal screens, so its equality to zero is natural. However, if you first integrate over the frequency and then take the limit, it will disperse. The conclusion is that all the formulas given with the real parts do not correspond to the results of (1) and (2), which are always valid and finite. These formulas can be obtained from the principle of argument [12] by summing the complex resonant frequencies of a finite structure (resonator) [12,17,18].

These frequencies lie in the upper half-plane of the complex plane $\omega$. After the transition to infinite transverse dimensions of the resonator, the sum turns into an integral of $\ln\left(f_e^{-1} + f_h^{-1}\right)$, where $f_{e,h} = 0$ are the characteristic (dispersion) equations of the structure. To obtain twice the sum of the real frequencies, Wick rotation is necessary: taking the integral along the right semicircle and the imaginary axis. In the right half-plane, the frequencies are complex conjugate. Multiplying them by $\hbar\coth(k\delta_T/2)/2$, we get free energy at a finite temperature. Differentiating it by *d*, we get (with a minus) the specific force.

It should be noted that during differentiation, the infinite contribution independent of *d* is eliminated, and the pressure is obtained as the difference between internal and external pressures. The cotangent poles lie on the vertical "imaginary" axis of the *k* plane. They correspond to half-quotes. For the Lifshitz structure $f_{e,h}(\xi,\kappa,d) = r_{e,h}^{-2}\exp(-2kpd) - 1$. At output (4), an imaginary frequency $\xi = i\omega$ is entered and substitutions $k_z = -i\sqrt{k^2+\kappa^2} = -ip'k$, $\coth(x) = i\cot(\hbar\xi/(2k_BT))$ are made. Finally (4) takes the form (2). However, in (4) the plane wave decomposition is clearly not visible, compared with formula (3). For (3) and (4), such a decomposition should express the fields in terms of the tensor Green function (GF) of the structure acting on the fluctuation current densities. Field correlations are expressed in terms of integrals of the components of GF and correlations of current densities, proportional to $\varepsilon''(\omega)$. As it is easy to see, the integral in (3) at $r_\mu^2 = 1$ also diverges. Indeed, the internal integral in the domain $k_0 < \kappa < \infty$ is imaginary (because $k_z = -i\tilde{k}_z$), and should be omitted. The integral over the domain $0 < \kappa < k_0$ contains the summand $-k_0^3/6$, i.e. the integral in (3) diverges in frequency when integrated. This divergence was noted by Lifshitz yet in [3]. His phrase is characteristic: "*Expression (2.2) is finite in itself, but it contains terms that diverge when integrated over d$\omega$. This is the term c $\omega^3$ that occurs when integrating dp terms in curly brackets.*" In [1,2], formulas are given without a full derivation, and there is no consistent conclusion of formula (3) in [3]. It is replaced by the phrase: "*After a number of transformations,* $F_\omega$ *can be represented in the following form*". Such a consistent conclusion should lead to the proportionality of all fields to dissipation $\varepsilon''(\omega)$ (according to formula (1.1) from [3]), which is included in the correlation. Therefore, the frequency integral for the force must include a quantity $\varepsilon''(\omega)$ as a multiplier, which is not present in (3) and (4). Thus, when using correlations for infinite space, Rytov's method allows us to approximately find only the temperature part of the force associated with dissipation. In particular, the interaction forces of CNTs [5,6] have been found, for which thermal interaction makes a significant contribution. For real structures, correlations depend on the shape of the bodies [7]. Formula (1.2) from [3] states that the delta correlations are correlated and proportional to $\varepsilon''(\omega)\Theta(\omega,T)$, where $\Theta(\omega,T)=\hbar\omega\coth(\hbar\omega/(2k_B T))/2$ is the average energy of the quantum oscillator. Such a correlation is possible only for an unlimited space [7] (for the derivation of formula (3), see [8]). For the structure under consideration, the correlation should depend on the coordinate, in particular on *d* and the boundary conditions. It is difficult to find such correlations: it is necessary to solve the problem of structure excitation by dipoles [7,9]. This means the influence of boundary conditions on the force. In this case, the Rytov's method using excitation by dipoles gives the correct result [7]. Further, in [3] the substitution $p=k_z=\sqrt{k_0^2-\kappa^2}$ is made, and the integral is transformed according to $pdp=\kappa d\kappa$. However, for the dependence indicated in [3], there should be $-pdp=\kappa d\kappa$, and the plus sign is obtained for the rotated plane *p*. In [3] it is stated that (2) turns into (1) at $T\to 0$. Let's check it out. We have $\delta_T\to\infty$, the points $k_n=2\pi n/\delta_T\to 0$ become closely spaced, and it is indeed possible to move from the sum to the integral. However, this is only possible for finite *d*, which corresponds to an infinitesimal contour integral. But, with integral (1), the contour integral also diverges. For small *d*, the result should depend on the ratio of *d* to the thermal length. For any case $d\to 0$, and $T\to 0$ the result (2) does not correspond to (1). Apparently, this is due to the fact that (2) takes into account the result from half-calculations, but does not take into account the contour integral, which is in (1). At high temperatures, the contribution of the residuals is large, since the dependence has an order $1/d^3$, and the contour integral can neglect. But at low *d* (and low temperature) it is essential. In [3], this explains the lack of contribution from the contour integral as follows: "integration along the sections of the imaginary axis between the poles gives purely imaginary values that fall out when taking Re." However, when using the argument principle, the function is considered analytical outside the poles, the real part is not taken, and the integral is equal to the sum of the residuals plus the contour integral. In this case, the integral over the infinite semicircle falls out, since the function there is zero.

Formula (1), as well as its generalizations to arbitrary plane–layered media, have been proved in a number of works (see, for example, [7, 10–16]) by several different methods: variational using Green's functions of electrodynamics [15], based on the principle of argument [10–14], using correlation relations based on Lorentz's lemma and excitation of structures by dipoles [7], using methods of quantum field theory in statistical physics [16]. In particular, when $r_{e,h}^2 = 1$ it follows from the result of Casimir $F/S = -P_C = \hbar c \pi^2 / (240 d^4)$, N/m$^2$ [17] for the attraction of ideally conducting plates in the far zone. In [18], based on the consideration of free energy by statistical mechanics methods in a resonant cavity with a finite temperature, the Casimir result, and its correction for low and high temperatures were obtained. It also indicates the corresponding errors in [3].

For metal plates described by PM of DP $\varepsilon(\omega) = \varepsilon_L - \omega_p^2/\omega^2$ with $\varepsilon_L = 1$ in a number of papers [2,3,15,19–23] the corrections to $P_C$ are given in the form $P = P_C (1 - c_1/(k_p d) + c_2/(k_p d)^2 ...)$ at $T$=0, where $a_1 = 16/3$ [15,19–23]. However, such PM is typical for a metal without impurities at zero temperature. For a non-zero temperature, temperature the corrections $P(T) = P_C (1 + a_1 \delta_T / d + a_2 \delta_T^2 / d^2 + ...)$, $\delta_T = \hbar c / (k_B T)$ are used. For real metals, the Lorentz term $\varepsilon_L$ is practically independent of frequency up to optical frequencies; it can have several Lorentz peaks in the UV range, and then tends to unity at very high frequencies. This resonant region has little effect on the force in the far zone, however, when numerically integrating over an infinite frequency limit, it is desirable to take it into account in order to obtain a force at short distances. Given the frequency–dependent Lorentz term $\varepsilon_L(\omega)$, it is more realistic for a metal to take DP in the form of $\varepsilon(\omega) = \varepsilon_L(\omega) - \omega_p^2 / (\omega^2 - i\omega\omega_c)$ [24], where $\omega_c$ is the collision frequency. The choice of DP metal model strongly influences the results. The absence of dissipation ($\omega_C$=0) means the absence of plasmon polaritons (PP), which make a significant contribution to the force at short distances. In this case, the field is completely determined by the emitted modes. In such PM, there is a second-order pole in DP at zero frequency.

It was shown in [25] that there is no continuous transition from DM to PM and a strong influence of dissipative PP at short distances was found, which is confirmed by experiment. PM does not provide such an impact. It was also found in [25] that due to strong nonlocality (long free path), the Casimir entropy approaches zero at $T \to 0$, when the polarization $s$ does not contribute to the classical part of the Casimir force. Also, in the near field, you cannot use the Leontovich impedance model, which was also often used to obtain corrections. Quite a lot of works have been devoted to the issue of choosing a DP metal model, for example, [21–29]. In fact, the Drude-Lorentz model is the most suitable for metal in electrodynamics [24]. For this model, there is a first-order pole at zero frequency, i.e. DP at low frequencies is the inverse of frequency. However, for pure metals, the CF behaves according to the law $\omega_c(T) = \omega_{c0}T/T_0$, where an CF is introduced at temperature $T_0$. Therefore, when $T \to 0$ the specified pole turns into a pole of the second order, and the result depends on the sequence of limits $T \to 0$ and $\omega \to 0$. When using DM, frequencies usually $\omega < \omega_c$ are used. It is of interest to consider significantly higher frequencies when using this model, while considering a temperature-dependent CF. Also, CF may depend on the frequency and decrease significantly with its increase. At room temperature $\omega_c \sim 3 \cdot 10^{13}$ Hz, and the length $\delta_c = 1/k_c \sim 10^{-5}$ m is comparable to $\delta_T = \hbar c/(k_B T)$. It is for such distances, at which the force is very small, that DM should be used. At short distances, the force should be estimated using at least a few Lorentz terms. However, at distances of the order of several nm, these terms are small, and DP is close to unity. Therefore, the largest error in calculating the force with such an approximation of DP should occur in the region of $10<d<100$ nm, where the contribution of the Lorentz terms is significant. As a rule, force measurements are not carried out at such distances.

Force measurements for metal samples are carried out at their final transverse dimensions $L$. The free electrons in them cannot move over distances greater than $L$. This means that there is a formal connection between them and a low frequency $\omega_{00} = c\pi\sqrt{2}/L$, similar to the coupling frequencies of electrons with an atom in the Lorentz term. This frequency corresponds to the lowest frequency of the conducting body, considered as a cluster (resonator). It cancels the DP pole at zero frequency, and for large $L$ becomes small but finite. Such a model is often called the Drude-Smith model (DSM) [30–32], and in it is desirable to consider the CF as a frequency—dependent value [31,32]. When measuring force for bodies with size $L \sim 10^{-4}$ m, this frequency is comparable to an CF.

There are a number of contradictory results in the literature on the correction of the Casimir force for metals (see, for example, [24,29]), which indicates the relevance of obtaining such corrections. In particular, DM leads to a violation of Nernst's thermal theorem [24–28], whereas it holds for PM, although DM is more realistic from a physical point of view. In [1], the coefficient $c_1$ is given with an error. It is convenient to designate $\delta_p = 1/k_p$, with $\delta_p$~120 nm (for silver). The indicated DP is characteristic of collision-free PM, which is the case for pure metals at zero temperature. It is more realistic to take into account the CF and take $\varepsilon(\omega) = \varepsilon_L(\omega) - \omega_p^2/(\omega^2 - i\omega\omega_c)$ where the Lorentz term $\varepsilon_L(\omega)$ is almost constant up to PF, and then has resonances and further tends to unity.

At low frequencies, the internal field is significantly affected, which can be taken into account, for example, by the Lorentz-Lorentz formula for dielectrics or Onsager for metals. However, this requires numerical calculations. For metals, the main contribution to strength occurs at $\omega << \omega_p$. In a number of papers [2,3,15,18,20,22,25], temperature corrections to $P_C$ are given. The methods of obtaining them for (3) are based on the Wick rotation $\omega = -ikc$, $k_z = -i\sqrt{\kappa^2 + k^2}$, $\coth(x) \rightarrow i\cot(z)$, $z = \hbar kc/(2k_B T)$, and on the Matsubara technique. As a result, there is a cotangent with poles on the imaginary axis, and the theory of residuals leads to the replacement of the integral by the sum of the Matsubara frequencies (2) [4]. In particular, in [2,3], by applying the Euler-Maclaurin formula, the or amendment $\Delta_T(d) = -(48/9)(d/\delta_T)^4$ or $\Delta_T(d) = -(16/3)(d/\delta_T)^4$, is given, and in [19] the sign of this amendment is reversed and the term $-240d/\delta_T$ is added. In [15], this correction is also given in the form $\Delta_T(d) = (16/3)(d/\delta_T)^4 - 240d/(\pi\delta_T)$, i.e., the result of [19] is confirmed. The temperature characteristic length $\delta_T = \hbar c/(k_B T)$ is introduced here. Thus, $x = \delta_T k_0 / 2$.

At low temperatures (low frequencies or high $d$) $x >> 1$, $\coth(x) \approx 1$, and the correction is small and should be considered at $\delta_p < d << \delta_T$. At room temperature $\delta_T \sim 2 \cdot 10^{-5}$ m, therefore, at shorter distances (usually $d/\delta_T << 1$), this correction is negative. In [18], the results with exponential multipliers without repulsive terms are presented, obtained by calculating the Casimir free energy taking temperature into account. In [20], this correction is given only in the form $\Delta_T(d) = (16/3)(d/\delta_T)^4$. The corrections in [2,3,15,19] are negative in this case, and always in [2,3], and in [15,19] up to the length $d = \delta_T \sqrt[3]{45/(4\pi)}$. According to these works at $d >> \delta_T$, the force generally ceases to depend on distance, and according to [2,3] it is repulsive, and according to [15,18,19] it is attractive. However, according to [2,3], the use of the correction is incorrect at $d >> \delta_T$. It makes sense to get amendments under different $d/\delta_T$. It should be noted that obtaining corrections is a complex procedure that strongly depends on the DP model [28] and the method of obtaining them. Thus, there is a discrepancy in the definition of amendments in the literature.

The aim of the work is a rigorous numerical determination of the force for thick metal plates in a wide range of ratios $d/\delta_c$, $d/\delta_T$ and $d/\delta_p$ for several DP models, taking into account the internal field and the Lorentz term, as well as obtaining analytical force corrections for the DP models considered. Although formulas (1) and (2) were published more than 70 years ago, there are no rigorous calculations based on them for several DP models and their comparison in the literature. This is especially true in the near field. The PM in the far zone $d/\delta_p > 1$ is mainly.

The Casimir result is obtained from (2) for $r_{e,h}^2 = 1$. For metals, this holds in the limit infinite $d$ if $\omega_{00} = 0$. When determining the amendments, we will further assume $r_{e,h}^2(\omega) = 1 - \alpha_{e,h}(\omega)$. Corrections in the far zone are determined at $\omega \to 0$. However, the value $\alpha_{e,h}(\omega) \to 0$ at is only for E-waves. For H-waves, we will limit the angles of incidence, since for sliding waves $r_{e,h}^2(\omega) \to 0$, and these areas do not contribute to the force. Provided that $r_{e,h}^2(\omega) \to 0$ at $\omega \to \infty$, and it additionally ensures the convergence of integrals of type (1) and (2).

The specific force can be represented in the form $P(d,T) = P_C(d)(1 - \Delta_0(d)) + \Delta_T(d,T)$ where the relative correction $\Delta_0(d)$ is determined by the deviations of the squares of RC $\alpha_{e,h}$ from unity at $T$=0, and the absolute correction $\Delta_T(d,T)$ gives the correction due to the finite temperature at infinite conductivity. We will get the temperature correction in several ways. The first method consists in using a very accurate decomposition of the hyperbolic cotangent $\coth(x) = 1 + \Delta(x)$ at low temperatures, where, $\Delta(x) \approx 2\exp(-2x)$ according to first approximation at large $x$ (high frequencies or low temperatures). High frequencies (low temperatures) correspond to low $d$. For low temperatures $k_0 >> 1/\delta_T$. When $k_0 = 1/\delta_T$ the adjusted for $\coth(x)$ less than one: $2/e < 1$. Large $d$ corresponds to small $x$ when $\coth(x)$ it is very different from unity, and the correction can no longer be small. Decomposition $\coth(x) = 1/x + x/3 - x^3/45....$ can be used at high temperatures. If we limit the spectral range to wavenumber $k_p = 2\pi/\lambda_p$, then low temperatures correspond $\delta_T >> \lambda_p$, i.e. $\delta_T > 1000$ nm. This gives the limits of applicability of the above formulas $130 < d < 1000$ nm for distances $d \sim \delta_T$ and less. To obtain an accurate result, use (2). Then $P_{e,h}(d,T) = P_{e,h}(d,0) + \Delta_{e,h}(d,T)$, $P(d,T) = P_e(d,T) + P_h(d,T)$.

The corrections are obtained based on formulas (1) and (2) using various DP models. Summarizing $\hbar\omega\coth(x)/2$ based on the principle of argument and using decomposition

$$\coth(x) = 1 + \Delta(x), \tag{5}$$

$$\begin{aligned}\coth(x) &= 1 + 2[\exp(-2x) + \exp(-4x) + ...] = \\ &= 1 + \Delta(x) = 1 + 2\exp(-2x)/(1 - \exp(-2x))\end{aligned}, \tag{6}$$

we see that on the imaginary axis $k$ the exponent has no poles, therefore . (7)

$$\Delta_T(d,T)=\frac{16\hbar c}{\pi^2\delta_T^4}\mathrm{Re}\int_{-i\pi}^{-i\infty}z^3\exp(-2z)\int_1^\infty\sum_{\mu=e,h}\frac{p^2dp}{r_\mu^{-2}(z,p)\exp(2izpd/\delta_T)-1}dz\,. \quad (7)$$

Decomposition (6) should be used at $x>>1$, otherwise the correction becomes large. To obtain small corrections, the lower limit in (7) is limited by the value $-i\pi$. At extremely high temperatures (large *d*), the decomposition $\coth(x)=1/x+x/3-x^3/45+....$ valid for small *x* ($x<\pi$) can be used. There are an infinite number of poles in $\coth(x)=i\cot(k\delta_T/2)$ on the imaginary frequency axis, half-inversions in which lead to formula (2). This formula, like (1), is a doubling of the integral along the imaginary axis by reducing it to the half-axis integral. The contour integral can be considered as

$$\tilde{F}'(d)=\frac{\hbar c}{2\pi^2}\,p.v.\int_0^\infty k^3\cot\left(\frac{k\delta_T}{2}\right)\int_1^\infty\sum_{\mu=e,h}\frac{r_\mu^2\exp(-2pkd)}{1-r_\mu^2\exp(-2pkd)}p^2dpdk\,. \quad (8)$$

The result (8) mast be taken with minus to (2) as $\coth(iz)=-i\coth(z)$. It is understood in the sense of the main meaning, i.e. contributions to the vicinity of the poles are eliminated. With an infinite imaginary upper limit, convergence in *p* is provided by the exponent in (8). Note that approaches using (7) and (8) have not been previously considered in the literature.

We will denote $\Delta_T(d,T)=\Delta_e(d,T)+\Delta_h(d,T)$. The squares of RC are complex and can be decomposed into series $r_\mu^2=1-a_1(p)k+a_2(p)k^2+...$ (for PM), and the integral over *p* is a complex function of *z*. Due to the presence of a rapidly oscillating exponent when integrated over p, the upper limit gives zero [3]. However, in reality, *z* runs through imaginary values, therefore, during the initial integration over *z*, the exponent is reset at the upper limit. The original integral (1) is valid, so the real part is taken.

## 2. Reflection from metal plate and force: different DP models

We consider the plates to be thick enough, i.e. we do not consider the waves reflected from the outer boundaries. Calculations show that it is sufficient to have a plate thickness of the order of tens of nm. Next, we consider the pressure as a sum $P(d)=P_e(d)+P_e(d)$. DP for a collision-free PM has the form $\varepsilon(k)=1+k_p^2/k^2$. In it $k_p=\omega_p c$, $k^2=-k_0^2$. It is possible for pure metals at zero temperature. The Drude-Lorenz model $\varepsilon(k)=\varepsilon_L(k)+k_p^2/(k^2+kk_c)$ takes into account CF $\omega_{c0}=k_c c$. The Lorentz term is almost constant up to PF, and tends to unity at $k\to\infty$. At low frequencies $\varepsilon(k)\approx\varepsilon_L(0)+k_p^2/(kk_c)$, where the first term can be neglected. For the Lorentz model with several resonant frequencies, we have

$$\varepsilon_L(\omega)=\varepsilon'(\omega)-i\varepsilon''(\omega)=1+\sum_{n=1}\frac{\omega_{pn}^2}{\omega_{0n}^2-\omega^2+i\omega\omega_{cn}}, \quad (9)$$

And for Drude-Lorentz model

$$\varepsilon(\omega)=\varepsilon'(\omega)-i\varepsilon''(\omega)=\varepsilon_L(\omega)+\frac{\omega_p^2}{i\omega\omega_{c0}-\omega^2}. \quad (10)$$

Once again, we note that for the selected dependence $\exp(i\omega t)$, all the poles of functions (9) and (10) are located in the upper half-plane of the complex frequency plane symmetrically relative to the imaginary axis, which ensures the principle of causality when using the Kramers-Kronig relations. In a dissipative environment, signals decay according to the law $\exp(i(\omega'+i\omega'')t)$. During Wick's rotation, the poles move into the right half-plane symmetrically relative to the real axis, i.e. they become complex conjugate, which, according to the principle of the argument, means the disappearance of imaginary parts in integrals (1), (2). For the Dude-Smith-Lorenz model, a small frequency square $\omega_{00}^2$ should be added to the denominator (10), which can be interpreted as spreading the sum in (9) from zero. Formulas (9) and (10) correspond to sparse media. In condensed media, different models can be used, for example, Onsager. If we use the Lorentz-Lorentz model

$$\frac{\varepsilon(\omega)-1}{\varepsilon(\omega)+2}=\frac{\chi(\omega)}{3},$$

Then for it the DP goes to infinity at $\chi(\omega)=3$, i.e. the medium must be sufficiently sparse. In the case of DP metal (10) at zero frequency $\varepsilon(0)=-2$, i.e. the model does not work for metals. For the Onsager model

$$\varepsilon(\omega)=1+\frac{3}{4}\left(\chi(\omega)-1+\sqrt{1+\frac{2}{3}\chi(\omega)+\chi^2(\omega)}\right) \quad (11)$$

at a low frequency and $\chi(\omega)\approx\omega_p^2/(i\omega\omega_{c0}-\omega^2)$, we have $\varepsilon(\omega)\approx 1/2+3\chi(\omega)/2$, i.e. the result is about one and a half times different from (10). Since the force is determined mainly by low frequencies, this is significant. If the measured samples are small in the transverse directions, the frequency $\omega_{00}$ can be comparable to an CF, and the measured force is significantly less. Turning to the imaginary frequency, we have the Onsager model

$$\chi(k)=\sum_{n=0}\frac{k_{pn}^2}{k_{0n}^2+k^2+kk_{cn}}, \quad (12)$$

$k_{p0}^2=k_p^2$. For non-conductive layers (plates), we simply set the square of PF to zero: $\omega_p^2=0$. At high frequencies, the value $\chi(k)$ is small, and according to the Lorentz-Lorentz model, we obtain $\varepsilon(k) \approx 1+\chi(k) \approx 1$, i.e., we have limiting values $1-\varepsilon'(\omega) \sim 1/\omega^2$ and $\varepsilon''(\omega) \sim 1/\omega^3$ at high frequencies. According to the Onsager model, we have $\varepsilon(\omega)=1+\chi(\omega)/2+3\chi^2(\omega)/8$, i.e. this model is more acceptable. Since at low frequencies the result of the model exceeds (10) by one and a half times, and at high frequencies it is two times less, it can be expected that both models (10) and (11) give similar results.

In the case of wavelengths of the order of the distances between atoms, the metal layer is no longer a continuous medium. Photoionization of atoms and plasma formation occur at such frequencies. Electrons oscillate around atoms with almost no collisions (dissipation), and the continuum model does not work. Interaction with the quantum vacuum takes place at the level of individual atoms and electrons. This justifies the introduction of clipping multipliers and frequency clipping in general.

We introduce normalized (multiplied by 120 Oh) wave conductivities $y_e(k_0,\kappa,\varepsilon)=k_0\varepsilon(k_0)/\sqrt{k_0^2\varepsilon(k_0)-\kappa^2}$ for E-waves and $y_h(k_0,\kappa,\varepsilon(k_0))=\sqrt{k_0^2\varepsilon(k_0)-\kappa^2}/k_0$ for H-waves. We take it in the vacuum gap as $y_{e,h}(k_0,1)$. The transition to the imaginary wavenumber gives $y_e(k,\kappa,\varepsilon)=k\varepsilon(k)/K_\varepsilon(\varepsilon(k),\kappa^2)$, $y_h(k,\kappa,\varepsilon(k))=K_\varepsilon(\varepsilon(k),\kappa^2)/k$, $K_\varepsilon(\varepsilon(k),\kappa^2)=\sqrt{k^2\varepsilon(k)+\kappa^2}$, $K_1(k,\kappa^2)=\sqrt{k^2+\kappa^2}$. The RC from the gap side is the essence $r_{e,h}(k,\kappa)=-(1-\alpha_{e,h})/(1+\alpha_{e,h})$, where

$$\alpha_{e,h}=y_{e,h}(k,\kappa,1)/y_{e,h}(k,\kappa,\varepsilon), \tag{13}$$

or $\alpha_e=K_\varepsilon/(\varepsilon K_1)$, $\alpha_h=K_1/K_\varepsilon$. It follows from (13) that for an infinitely large surface conductivity of a metal $r_{e,h}(k,\kappa)=-1$. The DP tends to infinity at $k\to 0$. Then $\varepsilon \approx \varepsilon_L+k_p^2/k_\varepsilon^2$ is large value, where the value $k_\varepsilon^2$ is small, and $k_\varepsilon^2=k^2$ for PM, $k_\varepsilon^2=k^2+kk_c$ for DM, and $k_\varepsilon^2=k_{00}^2+k^2+kk_c$ for DSM. The first term $\varepsilon_L$ is usually small compared to the second term, for metals at low frequencies $\varepsilon_L \sim 10$, but it is usually assumed that, generally speaking, it corresponds to infinitely large frequencies. For DM at low frequencies $\varepsilon=\varepsilon_L-(k_p/k_c)^2+k_p^2/kk_c$. For the DSM $k_{00}^2 << k_p^2$, the DP pole is missing. For PM $\alpha_e=0$, $\alpha_h=\kappa/\sqrt{k_p^2+\kappa^2}$, i.e. $r_e=-1$ at $k\to 0$. For it $r_h=-1$ only at $\kappa << k_p$, and it will be $\alpha_h=1$, $r_h=0$ for $\kappa >> k_p$. For DM at $k\to 0$, we have $\alpha_e=k_\varepsilon^2/k_p^2\to 0$, $\alpha_h=\kappa/\sqrt{k_p^2 k/k_c+\kappa^2}\to 1$, $r_h=0$. The difference between these models lies in the speed of marginal transitions. In order for the metal to reflect perfectly, it is necessary to fulfill the condition $k_p\to\infty$ for all models, i.e.

infinity DP at all frequencies. At an infinitely high frequency for all models $\varepsilon=1$, and $r_{e,h}=0$. Making the substitution of variables $\kappa^2=k^2\left(p^2-1\right)$, $\kappa d\kappa=k^2 pdp$, we have $K_1(k,p)=kp$, $K_\varepsilon(\varepsilon,k,p)=k\sqrt{\varepsilon(k)+p^2-1}$, and respectively

$$r_e(k,p)=-\frac{p\varepsilon(k)-\sqrt{\varepsilon(k)+p^2-1}}{p\varepsilon(k)+\sqrt{\varepsilon(k)+p^2-1}}, \tag{14}$$

$$r_h(k,p)=-\frac{\sqrt{\varepsilon(k)+p^2-1}-p}{\sqrt{\varepsilon(k)+p^2-1}+p}. \tag{15}$$

At $p$=1, both formulas give $r_{e,h}(k,1)=\left(1-\sqrt{\varepsilon(k)}\right)/\left(1+\sqrt{\varepsilon(k)}\right)$, which corresponds to the normal incidence of the waves. For $p\to\infty$ and finite $k$, the first formula gives $r_e(k,\infty)=(1-\varepsilon(k))/(1+\varepsilon(k))\to-1$, and the second $r_h(k,\infty)=0$. This is the case of large $\kappa$, i.e. zero sliding angle. These sliding angles correspond to the points responsible for the force in the near zone. Next, we consider the values $\alpha_{e,h}$ to be small. For $p\to\infty$ always $\alpha_h\to1$, therefore, it is necessary to limit $p$. To fulfill the condition of smallness $\alpha_h<1$, we require the fulfillment of the conditions $k<k_p$ and $p<p_{\max}=k_p/k_\varepsilon$. But $k_\varepsilon=k$ for PM, $k_\varepsilon=\sqrt{k^2+kk_c}$ for DM and $k_\varepsilon=\sqrt{k_{00}^2+k^2+kk_c}$ for DSM. When integrating over $p$, we will use this upper limit if the integrals diverge, otherwise we will take the upper limit to be infinite. The main problem is to approximate the root $s=\sqrt{\varepsilon(k)+p^2-1}$, because otherwise it is difficult to perform integration with respect to $p$. For all models, we have a high value of the root $s$:

$$s=\frac{\sqrt{k_p^2+k_\varepsilon^2 p^2}}{k_\varepsilon}\left(1+k_\varepsilon^2\frac{\varepsilon_L-1}{2\left(k_p^2+k_\varepsilon^2 p^2\right)}+...\right).$$

Substitutions should be made here

$$\sqrt{k_p^2+k_\varepsilon^2 p^2}=k_p\left(1+\frac{k_\varepsilon^2 p^2}{2k_p^2}+\frac{3}{8}\frac{k_\varepsilon^4 p^4}{k_p^4}+...\right),$$

$$\frac{1}{k_p^2+k_\varepsilon^2 p^2}=\frac{1}{k_p^2}-\frac{k_\varepsilon^2 p^2}{k_p^4}+\frac{k_\varepsilon^4 p^4}{k_p^6}-...$$

It is the limitation $p<k_p/k_\varepsilon$ that allows the use of such decompositions. Then we get

$$s=\sqrt{\varepsilon(k)+p^2-1}=\frac{k_p}{k_\varepsilon}+\frac{k_\varepsilon p^2}{2k_p}+\frac{3}{8}\frac{k_\varepsilon^3 p^4}{k_p^3}+$$

$$+\frac{(\varepsilon_L-1)}{4}\left(\frac{p^2k_\varepsilon^3}{k_p^3}-\frac{k_\varepsilon^5p^4}{k_p^5}+\frac{k_\varepsilon^7p^6}{k_p^7}-...\right)+. \tag{16}$$

$$+\frac{3(\varepsilon_L-1)}{16}\left(\frac{p^4k_\varepsilon^5}{k_p^5}-\frac{k_\varepsilon^7p^6}{k_p^7}+\frac{k_\varepsilon^9p^8}{k_p^9}-...\right)$$

In formulas (1) and (2) we have the decomposition

$$\sum_{\mu=e,h}\frac{p^2}{r_{e,h}^{-2}(k,p)\exp(2kpd)-1}=p^2\sum_{m=1}^{\infty}\sum_{\mu=e,h}r_\mu^{2m}(k,p)\exp(-2mkpd)\approx$$
$$\approx\sum_{m=1}^{\infty}\left[2p^2-\sum_{\mu=e,h}\left(m\alpha_\mu(k,p)p^2+\frac{m(m-1)}{2}\alpha_\mu^2(k,p)p^2\right)\right]\exp(-2mkpd). \tag{17}$$

Here $r_{e,h}^2=1-\alpha_{e,h}$, $r_{e,h}^{2m}\approx1-m\alpha_{e,h}+m(m-1)\alpha_{e,h}^2/2$, and

$$r_{e,h}^2(k,p)=\left(1-\frac{2\beta_{e,h}(k,p)}{1+\beta_{e,h}(k,p)}\right)^2, \tag{18}$$

$$\beta_e=\frac{\sqrt{\varepsilon(k)+p^2-1}}{p\varepsilon(k)}\approx\frac{\sqrt{\varepsilon(k)-1}}{\varepsilon(k)}\times$$
$$\times\left(\frac{1}{p}+\frac{p}{2(\varepsilon(k)-1)}+\frac{p^3}{8(\varepsilon(k)-1)^2}+...\right), \tag{19}$$

$$\beta_h=\frac{p}{\sqrt{\varepsilon(k)-1}\sqrt{1+p^2/(\varepsilon(k)-1)}}\approx$$
$$\approx\frac{1}{\sqrt{\varepsilon(k)-1}}\left(p-\frac{p^3}{2(\varepsilon(k)-1)}+\frac{3p^5}{8(\varepsilon(k)-1)^2}-...\right). \tag{20}$$

The value (19) is always less than one, and the value (20) is less than one at $p<p_{\max}=\sqrt{\varepsilon_L-1+k_p^2/k_\varepsilon^2}$ . It 's more convenient to take $p_{\max}=k_p/k_\varepsilon$ . We decompose in (18) by small quantities $\beta_{e,h}$ up to the sixth order: $\alpha_{e,h}\approx4\beta_{e,h}-8\beta_{e,h}^2+12\beta_{e,h}^3-16\beta_{e,h}^4+20\beta_{e,h}^5-24\beta_{e,h}^6$ . Next, we consider four orders of magnitude. For squares we get $\alpha_{e,h}^2\approx16\beta_{e,h}^2-64\beta_{e,h}^3+64\beta_{e,h}^4+...$. Now, for the square bracket in (17), up to the fourth order, we have

$$\sum_{e,h}p^2r_{e,h}^{2m}\approx2p^2-4mp^2B(k,p)+8m^2p^2C(k,p),$$

$$B(k,p)=\beta_e+\beta_h-5(\beta_e^3+\beta_h^3)+4(\beta_e^4+\beta_h^4), \tag{21}$$

$$C(k,p)=\beta_e^2+\beta_h^2-4(\beta_e^3+\beta_h^3)+4(\beta_e^4+\beta_h^4), \tag{22}$$

and the coefficients in these formulas are

$$\beta_e^2 \approx \frac{\varepsilon(k)-1}{\varepsilon^2(k)}\left(\frac{1}{p^2}+\frac{1}{\varepsilon(k)-1}\right),$$

$$\beta_e^3 \approx \frac{(\varepsilon(k)-1)^{3/2}}{\varepsilon^3(k)}\left(\frac{1}{p^3}+\frac{3}{2p(\varepsilon(k)-1)}\right),$$

$$\beta_e^4 \approx \frac{(\varepsilon(k)-1)^2}{\varepsilon^4(k)}\left(\frac{1}{p^4}+\frac{2}{p^2(\varepsilon(k)-1)}\right),$$

$$\beta_h^2 \approx \frac{1}{\varepsilon(k)-1}\left(p^2-\frac{p^4}{\varepsilon(k)-1}\right), \tag{23}$$

$$\beta_h^3 \approx \frac{1}{(\varepsilon(k)-1)^{3/2}}\left(p^3-\frac{3p^5}{2(\varepsilon(k)-1)}\right),$$

$$\beta_h^4 \approx \frac{1}{(\varepsilon(k)-1)^2}\left(p^4-\frac{2p^6}{\varepsilon(k)-1}\right).$$

We also have the coefficients

$$\frac{\sqrt{\varepsilon(k)-1}}{\varepsilon(k)} = \frac{k_\varepsilon}{k_p}\left(1+k_\varepsilon^2\frac{\varepsilon_L-1}{2k_p^2}+\ldots\right)\left(1-\frac{k_\varepsilon^2\varepsilon_L}{k_p^2}+\ldots\right)= \\ = \frac{k_\varepsilon}{k_p}-\frac{k_\varepsilon^3(\varepsilon_L+1)}{2k_p^3}-\frac{k_\varepsilon^5\varepsilon_L(\varepsilon_L-1)}{2k_p^5}+\ldots,$$

$$\frac{\varepsilon(k)-1}{\varepsilon^2(k)} = \frac{k_\varepsilon^2}{k_p^2}-\frac{k_\varepsilon^4(\varepsilon_L+1)}{k_p^4}+\frac{k_\varepsilon^6(\varepsilon_L+1)^2}{4k_p^6}+\ldots$$

$$\frac{(\varepsilon(k)-1)^{3/2}}{\varepsilon^3(k)} = \frac{k_\varepsilon^3}{k_p^3}-3\frac{k_\varepsilon^5(\varepsilon_L+1)}{2k_p^5}+\ldots,$$

$$\frac{(\varepsilon(k)-1)^2}{\varepsilon^4(k)} = \frac{k_\varepsilon^4}{k_p^4}-4\frac{k_\varepsilon^6(\varepsilon_L+1)}{2k_p^6}+\ldots, \tag{24}$$

$$\frac{1}{\sqrt{\varepsilon(k)-1}} = \frac{k_\varepsilon}{k_p}-k_\varepsilon^3\frac{\varepsilon_L-1}{2k_p^3}+\ldots,$$

$$\frac{1}{\varepsilon(k)-1} = \frac{k_\varepsilon^2}{k_p^2}-k_\varepsilon^4\frac{\varepsilon_L-1}{k_p^4}+\ldots.,$$

$$\frac{1}{(\varepsilon(k)-1)^{3/2}} = \frac{k_\varepsilon^3}{k_p^3}-3k_\varepsilon^5\frac{\varepsilon_L-1}{2k_p^5}+\ldots,$$

$$\frac{1}{(\varepsilon(k)-1)^2} = \frac{k_\varepsilon^4}{k_p^4}-2k_\varepsilon^6\frac{\varepsilon_L-1}{k_p^6}+\ldots.$$

These formulas are valid for all DP models. For PM with $\varepsilon_L \neq 1$, formulas for $B(k,p)$ and $C(k,p)$ up to the fifth order are given in Appendix (A1), (A2).

When calculating Casimir-Lifshitz forces, integrals often arise, given in Appendix, formulas (A3)–(A8). When integrating (17), it is necessary to calculate integrals from $p^l k^{3+\nu}\exp(-2mkpd)dpdk$. Integrals of non-negative degrees *l* are determined by the formula, $l=\mu-2-\nu\geq 0$,

$$\varphi_l(\mu,\nu,k,p_{\max},d)=\int_1^{p_{\max}} p^{\mu+2}k^{3+\nu}\exp(-2mkpd)dpdk=\frac{\Gamma(4+\nu)}{(2md)^{4+\nu}}\times$$
$$\times\left[\frac{p^l\exp(-2mpd)}{(-2md)}-\frac{lp^{l-1}\exp(-2mpd)}{(-2md)^2}+...+(-1)^l\frac{l!\exp(-2mpd)}{(-2md)^{l+1}}\right]_{p=1}^{p=p_{\max}}. \quad (25)$$

Integrals of negative values are determined by integration in parts $(l=\mu+2+\nu\geq 0)$:

$$\psi_l(\mu,\nu,k,p_{\max},d)=\int_0^{\infty}\int_1^{p_{\max}} k^3\frac{\exp(-2mkpd)}{p^{\mu-2}}dpdk=\frac{\Gamma(4+\nu)}{(2md)^{4+\nu}}\times$$
$$\times\left[\frac{\exp(-2mpd)}{p^l(-2md)}+l\frac{\exp(-2mpd)}{p^{l+1}(-2md)^2}+l(l-1)\frac{\exp(-2mpd)}{p^{l+1}(-2md)^3}+...\right]_{p=1}^{p=p_{mzx}}. \quad (26)$$

Obviously, the terms of the series in (26) decrease, so it is convenient to leave three terms to achieve the stated degree of expansion. The final upper limit $p_{\max}=k_p d$ should be, generally speaking, in the decompositions of magnetic quantities, and for electrical ones it is possible to be infinite, which simplifies the formulas. A strictly finite limit is needed to use root decompositions. However, large *p* corresponds to PP, which do not contribute to the force in the far zone, and it is quite possible to take an infinite limit for decompositions of magnetic quantities if the integrals converge, since then we omit infinitesimally small quantities $p^l_{\max}\exp(-2mp_{\max}d)/(2md)$ for large *d*, as well as even smaller quantities. Omitting such corrections does not change the asymptotic expansions in negative powers of *d*. Assuming $p_{\max}=\infty$, we introduce notations $\varphi_l$ for (25) and $\psi_l$ for (26). In (2), we should integrate only with respect to *p*. The dependence of these values on *m* is not shown for brevity. For electrical quantities with a multiplier *m*, if we take into account terms up to the fourth order in , we have the values *l*=-2, -1, 0, 1, 2, 3, and for magnetic quantities, respectively, *l*=3,4,5,6,7,8. For electrical quantities with a multiplier $m^2$, we have the values *l*= -2,-1,0,1,2, and for magnetic quantities *l*=4,5,6,7,8. After integration in (17), we obtain

$$\sum_{m=1}^{\infty}\left[2A(m,d)-4mB(m,d)+8m^2C(m,d)\right], \quad (27)$$

where $A(m)=6/(2md)^4$, and coefficients (21), (22) are determined by formulas (23) – (26). These coefficients include values (24) with coefficients $(k_\varepsilon/k_p)^\nu$, $\nu=1,2,3,...$ .. When integrating them over *k* with a multiplier $k^3$ and taking into account the fact that for PM $k_\varepsilon = k$, and for DM $k_\varepsilon = \left(k^2 + kk_c\right)^{1/2}$, we obtain $\Gamma(4+\nu)/\left[(2md)^{4+\nu} k_p^\nu\right]$ for PM, and a rather complex formula for DM [33]:

$$\int_0^\infty \frac{k^{3+\nu/2}}{k_p^\nu}\left(k + k_c\right)^{\nu/2} \exp(-2mkd)dk = \\ = \frac{\exp(2mk_c d)\Gamma(4+\nu/2)k_c^{4+\nu}}{k_p^\nu}\Psi(4+\nu/2, 5+\nu, 2mk_c d).$$

Here $\Psi$ is a degenerate hypergeometric function. In the small *k* approximation, we have the decomposition

$$\left(k + k_c\right)^{\nu/2} = k_c^{\nu/2}\left(1 + \frac{\nu k}{2k_c} + \frac{\nu}{4}\left(\frac{\nu}{2} - 1\right)\left(\frac{k}{k_c}\right)^2 + \ldots\right),$$

which, after integration and summation over *m*, gives terms proportional to $\left(k_c d\right)^{-(4+\nu/2+l)}$, where *l*=0,1,2.... Such decompositions should be considered at distances $d >> 1/k_c$ significantly greater than plasma length $\delta_p = 1/k_p$. Moreover, for odd $\nu$ we obtain half-integer powers in the expansions. For PM, we obtain integer negative degrees of decomposition, valid for $d >> 1/k_p$. Therefore, DM should be used at distances significantly greater than $10^{-5}$ m. You can also provide formulas using special functions for the DSM. Integration over *k* with a multiplier $k^3$ for $2A(m,k)$, followed by summation, gives the value $d^{-4}\Gamma(4)\varsigma(4)/8$, i.e. taking into account the multiplier in (1) and $\xi = ck$, we obtain the Casimir force. The following terms in expansion (25), after integration and summation, give power expansions $d^{-\nu}$, $\nu$>4, i.e. corrections. The coefficients (22) are greatly simplified if we put $\varepsilon_L = 1$. This is used for all known small - magnitude decompositions of force in the literature, and PM c is used in all sources. In [21], such a decomposition is obtained up to the fourth order, and in [22] up to the sixth order. However, in these works, decomposition is used by parameter $\alpha = k/k_p = (x/p)\delta_p/(2d)$ that is small, as well as by a parameter $A = \exp(x)/(\exp(x) - 1)$ that is not small. Moreover, it tends to infinity at $x \to 0$. In addition, the magnitude $\varepsilon_L \neq 1$ significantly affects the strength for real metals. Therefore, these results need to be clarified. The above formulas allow us to obtain a decomposition up to and including the sixth order for all DP models. For DM $\varepsilon = \varepsilon_L + k_p^2/\alpha$, and decomposition by a small parameter $\alpha = k_c k + k^2$ leads to more complex results. However, obtaining high-order corrections does not make significant sense compared to numerically determining the integrals in (1) and (2).

Let's note the following interesting facts. The main term (the Casimir force) at $T$=0 is obtained regardless of the order of integration, first in $p$ and then in $k$, or vice versa. The first case should be used in (2), when summation takes place instead of integration over $k$. As can be seen above, in this case, divergences do not arise with an infinite upper limit. In formula (1), it is convenient to use decomposition (17) and make a substitution (as in [12,21]) $k = y/(2pd)$, $dk = dy/(2pd)$, integrating first over $y$. Then, in the external integral over $p$, instead of the multiplier $p^2$, the multiplier $1/p^2$ appears. In this case, for convergence in the decompositions of the magnetic modes, negative powers of $p$ less than -1 must occur. This is the case for PM $k_\varepsilon = y/(2pd)$, but it is not the case for DM, since $k_\varepsilon \approx \sqrt{yk_c/(2pd)}$ and there is no compensation of positive powers of $p$ in formulas (21) either. The result for PM does not depend on the order of integration. However, for DM, DSM, and the Drude-Lorentz model, the terms with $p \geq 1$ arise in the coefficients $\beta_h$. When integrating first over $p$ and then over y, divergences are obtained. When integrating first over $y$ and then over $p$, in order to obtain the final results in a number of terms, a finite upper limit $p_{\max}$ should be taken. Since the main contribution to the force is given by the frequencies $\omega \sim c/d$ or $k \sim 1/d$, for PM $p_{\max} = k_p d$, and for DM $p_{\max} = k_p d^{1/2}/k_c^{1/2}$. In fact, this means ignoring the mods for which $r_h \approx 0$. In this case, terms with reduced values of negative degrees of $d$ arise. For PM, it is also easiest to integrate first by y and then by $p$.

Thus, it is possible to limit frequencies to $k_p$ or $k_c$ values, which fully corresponds to the fact that high frequencies do not contribute to the force at $d > \delta_p$ or $d > \delta_c$. This leads to complex formulas containing small exponentials, which, in the limit of large $d$, go into the ones given below. If the frequencies are not limited, then for DM and DSM, $p$ should be limited in some terms of the magnetic mode expansions. This corresponds to the fact that these modes do not contribute to the force at high $p$. However, it is possible not to decompose the roots at all, but to calculate integrals without using decompositions. Next, we will demonstrate this by calculating the integral over $p$ in parts. In the case of decompositions for (1), we obtain

$$F(d)/S = \frac{\hbar c\pi^2}{240d^4}\left(1 - c_1\frac{\delta}{d} + c_2\left(\frac{\delta}{d}\right)^2 - c_3\left(\frac{\delta}{d}\right)^3 + c_4\left(\frac{\delta}{d}\right)^4 + ...\right). \quad (27)$$

It is not difficult to verify that from the above relations for PM with $\varepsilon_L \neq 1$ we obtain $\delta = \delta_p = 1/k_p$, $c_1 = 16/3$, $c_2 = 24$, i.e. the same coefficients as in [20,21], and the first coefficient corresponds to earlier works [15,19,20]. However, the following two coefficients already depend on $\varepsilon_L$. Exactly,

$$c_3 = \frac{10\varsigma(6)}{\varsigma(4)}(2\varepsilon_L - 1) + \frac{1950\varsigma(5) - 1200\varsigma(6)}{7\varsigma(4)}, \tag{28}$$

$$c_4 = \frac{1400(2\varsigma(6) - \varsigma(7))/3 - 112\varsigma(6)(\varepsilon_L + 1)}{\varsigma(4)}. \tag{29}$$

Formulas up to the fourth order of accuracy were used to obtain (28) and (29):

$$\beta_e = \frac{k}{k_p p} + \frac{k^3}{2k_p^3}\left(p - \frac{(\varepsilon_L + 1)}{p}\right),$$

$$\beta_h = \frac{k}{k_p} p - \frac{k^3}{k_p^3}\left(\frac{p^3}{2} + p\frac{\varepsilon_L - 1}{2}\right),$$

the degrees of which are also taken up to the 4th order by $y = k/(2pd)$. For DM $\delta = \delta_c = 1/k_c$, $\varepsilon(k) = \tilde{\varepsilon} + k_p^2/(kk_c)$, $\tilde{\varepsilon} = \varepsilon_L - k_p^2/k_c^2$ at $k < k_c$, and the coefficients take the form

$$\beta_e = \frac{(kk_c)^{1/2}}{pk_p} + \frac{(kk_c)^{3/2}}{2k_p^3}\left(p - \frac{(\tilde{\varepsilon} + 1)}{p}\right) - \frac{p(kk_c)^{5/2}}{4k_p^5} 2\varepsilon_L + \ldots, \tag{30}$$

$$\beta_h = \frac{(kk_c)^{1/2}}{k_p} p - \frac{(kk_c)^{3/2}}{2k_p^3}(p^3 + p) + \frac{(kk_c)^{5/2}}{8k_p^5}\left(3p^5 + 6p^3(\tilde{\varepsilon} - 1)\right) + \ldots. \tag{31}$$

Omitting the third term in (30), (31), we have

$$\beta_e = \frac{(yk_c)^{1/2}}{(2d)^{1/2} p^{3/2} k_p} + \frac{(yk_c)^{3/2}}{2(2d)^{3/2} k_p^3}\left(\frac{1}{p^{1/2}} - \frac{(\tilde{\varepsilon} + 1)}{p^{5/2}}\right),$$

$$\beta_e^2 = \frac{yk_c}{2dk_p^2 p^3} + \frac{(yk_c)^2}{(2d)^2 k_p^4}\left(\frac{1}{p^2} - \frac{(\tilde{\varepsilon} + 1)}{p^4}\right),$$

$$\beta_e^3 = \frac{(yk_c)^{3/2}}{(2d)^{3/2} p^{7/2} k_p^3},$$

$$\beta_e^4 = \frac{(yk_c)^2}{(2d)^2 k_p^4 p^6},$$

$$\beta_h = \frac{(yk_c)^{1/2}}{(2d)^{1/2} k_p} p^{1/2} - \frac{(yk_c)^{3/2}}{2(2d)^{3/2} k_p^3}\left(p^{3/2} + \frac{1}{p^{1/2}}\right),$$

$$\beta_h^2 = \frac{(yk_c)}{(2d)k_p^2} p - \frac{(yk_c)^2}{(2d)^2 k_p^4}\left(p^{3/2} + \frac{1}{p^{1/2}}\right),$$

$$\beta_h^3 = \frac{(yk_c)^{3/2}}{(2d)^{3/2} k_p^3} p^{3/2} - \frac{(yk_c)^{5/2}}{2(2d)^{5/2} k_p^5}\left(p^{5/2} + p^{1/2}\right),$$

$$\beta_h^4 = \frac{(yk_c)^2}{(2d)^2 k_p^4} p^2.$$

After integration over *y*, and then over *p*, taking into account $p_{\max} = k_p d^{1/2} / k_c^{1/2}$ and subsequent summation, we obtain a decomposition with fractional indices.

$$F(d)/S = \frac{\hbar c \pi^2}{240 d^4}\left(1 + \frac{c_{1/2} k_c^{1/2}}{k_p d^{1/2}} + \frac{c_1 k_c [\ln(p_{\max}) + 1/4]}{k_p^2 d} + \frac{c_{5/4} k_c^{5/4}}{k_p^{5/2} d^{5/4}} + \right.$$
$$\left. + \frac{c_{3/2} k_c^{3/2}}{k_p^3 d^{3/2}} + \frac{c_{7/4} k_c^{7/4}}{k_p^{7/2} d^{7/4}} + \frac{c_2 k_c^2}{k_p^4 d^2} + \frac{c_{9/4} k_c^{9/4}}{k_p^{9/2} d^{9/4}} + \frac{c_{5/2} k_c^{5/2}}{k_p^5 d^{5/2}} + \frac{c_{11/4} k_c^{11/4}}{k_p^{11/2} d^{11/4}} + \ldots \right), \tag{32}$$

$$c_{1/2} = -2^{9/2} \frac{3\Gamma(4+1/2)\varsigma(3+1/2)}{5\Gamma(4)\varsigma(4)},$$

$$c_1 = \frac{\varsigma(3)}{2\varsigma(4)},$$

$$c_{5/4} = \frac{\Gamma(5+1/2)(21\varsigma(4+1/2) - 32\varsigma(3+1/2))}{2^{3/2}\Gamma(4)\varsigma(4)},$$

$$c_{3/2} = \Gamma\left(5+\frac{1}{2}\right)\frac{4\varsigma\left(4+\frac{1}{2}\right)\left(\frac{\tilde{\varepsilon}+1}{7} - 1\right) + \frac{16}{9}\left[32\varsigma\left(3+\frac{1}{2}\right) - 20\varsigma\left(4+\frac{1}{2}\right)\right]}{2^{5/2}\Gamma(4)\varsigma(4)},$$

$$c_{7/4} = \frac{1}{\Gamma(4)\varsigma(4)}\left(\Gamma(6+1/2)\frac{8\varsigma(4+1/2) - 5\varsigma(5+1/2)}{2^{3/2}3} - 2\Gamma(6)\varsigma(4)\right),$$

$$c_2 = \frac{\Gamma(6)}{\Gamma(4)\varsigma(4)}\left(\varsigma(4)\left(\frac{5}{3} - \frac{(\tilde{\varepsilon}+1)}{5}\right) + \frac{12\varsigma(5)}{7} - \frac{96\varsigma(4)}{7}\right).$$

## 3. Low temperature corrections

First, let's consider correction (7), which we will write down in approximate form

$$\Delta_T(d,T) = \frac{16\hbar c}{\pi^2 \delta_T^4} \mathrm{Re} \int_{-i\pi}^{-i\infty} z^3 \exp(-2z) \int_1^{\infty} \sum_{\mu=e,h} \frac{p^2 dp}{r_\mu^{-2}(z,p)\exp(2izpd/\delta_T) - 1} dz. \tag{33}$$

The series expansion is legal, since $\left|r_{e,h}^2\right| < 1$, and for imaginary *z*, the exponent decays. For simplicity, we use decompositions of RC squares up to the second order by $\beta_{e,h}$: $r_{e,h}^{2m} \approx 1 - 4m\beta_{e,h} + 8m^2\beta_{e,h}^2$. For PM we have

$$p^2\left[r_e^{2m}(z) + r_h^{2m}(z)\right] = 2p^2 - 4mi\left(\frac{z}{k_p\delta_T}(p+p^3) - \frac{z^3}{2k_p^3\delta_T^3}\left(p^3(2-\varepsilon_L) - p^5 - \frac{(\varepsilon_L+1)}{p}\right)\right) +$$
$$+ 8m^2\left(\frac{-z^2}{k_p^2\delta_T^2}(1+p^4) + \frac{z^4}{k_p^4\delta_T^4}\left(p^3 - \frac{(\varepsilon_L+1)}{p} + p^6 - p^4(\varepsilon_L+1)\right)\right). \tag{34}$$

When $r_{e,h}^2 = 1$, after calculating the integral over $p$ in (33) at $\Delta(z) = 2\exp(-2z)$, we have the function

$$\varphi_{0m}(z) = \left(-i\delta_m z^2 - 2z\delta_m^2 + 2i\delta_m^3\right)\exp\left(-\left(i/\delta_m + 2\right)z\right).$$

It indicates a dimensionless parameter $\delta_m = \delta_T/(2md)$. Consider the functions $\varphi_{0ml}(z) = z^l \exp(-(iz/\delta_m + 2))$ . They fade out on the imaginary axis. After integrating them over $z$, we obtain for $l$=2,1,0

$$\Phi_{0m2}(\delta_m) = \left[\frac{\pi^2}{(2 + i/\delta_m)} + \frac{2i\pi}{(2 + i/\delta_m)^2} - \frac{2}{(2 + i/\delta_m)^3}\right]\exp\left(-\frac{\pi}{\delta_m}\right) ,$$

$$\Phi_{0m1}(\delta_m) = \left[\frac{i\pi}{(2 + i/\delta_m)} - \frac{1}{(2 + i/\delta_m)^2}\right]\exp\left(-\frac{\pi}{\delta_m}\right),$$

$$\Phi_{0m0}(\delta_m) = -\frac{1}{(2 + i/\delta_m)}\exp\left(-\frac{\pi}{\delta_m}\right).$$

After integration $\varphi_{0m}(z)$, we get

$$\Phi_{0m}(\delta_m) = -i\delta_m\Phi_{0m2}(\delta_m) - 2\delta_m^2\Phi_{0m1}(\delta_m) + 2i\delta_m^3\Phi_{0m0}(\delta_m).$$

The parameter $\delta_m$ is small at $d > \delta_T$. Then, to a first approximation, the formulas can neglect the twos, $\Phi_{0m2}(\delta_m) \approx -2i\delta_m^3 \exp(-\pi/\delta_m)$ , $\Phi_{0m2}(\delta_m) \approx -2i\delta_m^3 \exp(-\pi/\delta_m)$, and $\Phi_{0m}(\delta_m) = -6\delta_m^4 \exp(-\pi/\delta_m)$, i.e., taking the real part in (33) was not necessary. We get

$$\Delta_T(d,T) \approx -\frac{6\hbar c}{\pi^2 d^4}\sum_{m=1}^{\infty}\frac{1}{m^4}\exp\left(-\frac{2m\pi d}{\delta_T}\right), \tag{35}$$

$$\tilde{F}(d,T) \approx \tilde{F}_C\left(1 - \frac{1440}{\pi^4}\sum_{m=1}^{\infty}\frac{1}{m^4}\exp\left(-\frac{2m\pi d}{\delta_T}\right)\right). \tag{36}$$

Here the exponent is very small, therefore $\tilde{F}(d,T) \approx \tilde{F}_C(1 - 13.6\exp(-2\pi d/\delta_T))$, i.e. low temperature slightly reduces the force in the area $d > \delta_T$. Already if $d = \delta_T$, we get a decrease of about 2.5%. In another limiting case $d << \delta_T$, the parameter is large, $\Phi_{0m2}(\delta_m) \approx \pi^2/2 - 1/4 + i\pi/4$, $\Phi_{0m1}(\delta_m) \approx i\pi/2 - 1/4$, $\Phi_{0m0}(\delta_m) = -1/2$ , and we obtain $\mathrm{Re}\,\Phi_{0m}(\delta_m) \approx \delta_m^2 \exp(-\pi/\delta_m)/2$, and

$$\Delta_T(d,T) = \frac{32\hbar c d^2}{\pi^2\delta_T^6}\sum_{m=1}^{\infty} m^2 \exp\left(-\frac{2m\pi d}{\delta_T}\right). \tag{37}$$

In the limit, replacing the sum with an integral, we have

$$\sum_{m=1}^{\infty} m^2 \exp\left(-\frac{2m\pi d}{\delta_T}\right) = 2\left(\frac{\delta_T}{2\pi d}\right)^3 + $$
$$+2\left(\frac{\delta_T}{2\pi d}\right)^2 + \left(\frac{\delta_T}{2\pi d}\right) \approx 2\left(\frac{\delta_T}{2\pi d}\right)^3 ,$$

$\Delta_T(d,T) = 8\hbar c/\left(\pi^4 \delta_T^3 d\right)$, and $\tilde{F}(d,T) \approx \tilde{F}_C\left(1+\left(1920/\pi^6\right)d^3/\delta_T^3\right)$. It is easy to see that taking into account the discarded terms yields the following terms of the decomposition with $d^4/\delta_T^4 + \ldots$.

For a real metal, the function $\varphi_m(z)$ should be obtained by integrating (34). The value $\varepsilon_L$ affects only after taking into account the 3rd and higher orders of magnitude in *z*. Let's give the influence of the first order. In this case $\varphi_m(z) = \varphi_{0m}(z) - \varphi_{1m}(z)$, and

$$\varphi_{1m}(z) = \frac{4m}{k_p \delta_T} \exp\left(-\frac{iz}{\delta_m}\right)\left(\frac{-\left(p+p^3\right)z^3}{\left(p/\delta_m\right)} + \frac{4iz^2}{\left(p/\delta_m\right)^2} + \frac{6z}{\left(p/\delta_m\right)^3} - \frac{6i}{\left(p/\delta_m\right)^4}\right).$$

Integrating this function over *z*, we obtain the function $\Phi_{1m}$ and see that, in addition to decompositions of type (35), it also gives a multiplier $\delta_p/d$ . Thus, in the decomposition of type (36), (37) for a real metal, the terms with $\left(\delta_p/d\right)^\nu$ will also appear. Formally, the decompositions obtained in section 2 can be used. In order not to overload the work with formulas, we do not present these results. Further, we obtain similar expansions in the form of double series from formula (2), as well as from formula (8).

Let us consider the integration of (33) first with respect to *z*, and then with respect to *p* for . Let's denote $b_m = 2mid/\delta_T$, $a = 2$, $a$=0. The result of integrating the function $p^2 z^3 \exp\left(-\left(a+b_m p\right)z\right)$ for $a$=2 and $a$=0 is the function

$$\psi_m(a,p) = -p^2 \exp\left(-i\pi b_m p\right)\left[\frac{-i\pi^3}{\left(a+b_m p\right)} + \frac{3\pi^2}{\left(a+b_m p\right)^2} - \frac{6i\pi}{\left(a+b_m p\right)^3} - \frac{6}{\left(a+b_m p\right)^4}\right].$$

Since the parameter $b_m$ is small for small indexes, the main contribution is provided by the function $\psi_m(0,p)$, and for small $p << 1/\left(\pi b_m\right)$ it is possible to take $\exp\left(-i\pi b_m p\right) \approx 1$. Then $\psi_m(0,p) \approx 6\delta_m^4\left[1 - i\pi^3/2\delta_m + 3\pi^2/\delta_m^2\right]$, i.e., we get the same result, but in a more complicated way. At the same time, since there is still a decaying exponent, we zeroed the substitution at the upper limit $p = \infty$. In general, formulas (92.1) – (92.4) from [34] should be used to calculate the integrals.

Since the corrections at small $\delta_T$ are not small ones, the obtaining accurate ratios based on (33) is problematic. Moreover, this applies to real metals. In [18], corrections for small and large $\delta_T/d$ were obtained close to the Casimir method by calculating the free energy in a resonator with equilibrium thermal radiation and ideally conducting walls separated by an ideally conducting shield. The Euler-Maclaurin formula is used. At low temperatures ($d/\delta_T << 1$), small parameter decompositions are used, leading to the result

$$\tilde{F}(d,\delta_T) = \tilde{F}_C(d)\left[1 - \frac{240}{\pi}\frac{d}{\delta_T}\exp(-\pi\delta_T/d) + \frac{16}{3}\left(\frac{d}{\delta_T}\right)^4\right], \quad (38)$$

and at high temperatures ($d/\delta_T >> 1$), the Poisson summation formula is used, giving the result

$$\tilde{F}(d,\delta_T) = \frac{\hbar c}{4\pi d^3\delta_T}\left[\varsigma(3) + \left(2 + 8\pi\frac{d}{\delta_T} + \left(4\pi\frac{d}{\delta_T}\right)^2\right)\exp\left(-\frac{4\pi d}{\delta_T}\right)\right]. \quad (39)$$

Assuming in (2) first $r_{n\mu}^{-2} = 1$ and isolating the term with $n$=0, we obtain

$$\tilde{F}(d,T) = \frac{k_B T}{\pi}\sum_{n=0}^{\infty}\frac{k_n^3}{1+\delta_{n0}}\int_1^{\infty}\frac{p^2 dp}{\exp(2pk_n d)-1} =$$
$$= \frac{\hbar c}{4\pi d^3\delta_T}\left[\varsigma(3) + \sum_{m=1}^{\infty}\sum_{n=1}^{\infty}\exp\left(-\frac{4\pi nmd}{\delta_T}\right)\left(\frac{16\pi^2 n^2 d^2}{m\delta_T^2} + \frac{8\pi nd}{m^2\delta_T} + \frac{2}{m^3}\right)\right]. \quad (40)$$

At extremely long distances $F = k_B T\varsigma(3)/(4\pi d^3)$. In the case of large $d/\delta_T$, leave the first term in parentheses and take the first terms in sums. Then

$$\tilde{F} = \frac{\hbar c}{4\pi d^3\delta_T}\left[\varsigma(3) + \frac{16\pi^2 d^2}{\delta_T^2}\exp\left(-\frac{4\pi d}{\delta_T}\right)\right]. \quad (41)$$

In the far zone, it is sufficient to take into account the first term in this formula, i.e. we have a well-known result $\tilde{F}(d,T) \approx k_B T\varsigma(3)/(4\pi d^3)$ [25]. In the case of extremely small $d/\delta_T$, you should use many terms in the sums. Ignoring the first two terms in parenthesis, we get

$$\tilde{F} \approx \frac{k_B T}{4\pi d^3}\left(\varsigma(3) + 2\sum_{\substack{n=1\\m=1}}^{\infty}\frac{1}{m^3}\exp(-4\pi mnd/\delta_T)\right).$$

Moving from the sum of $n$ to the integral, i.e. taking $dn = dz\delta_T/(4\pi md)$, we get

$$\tilde{F} \approx \frac{k_B T}{4\pi d^3}\left(\varsigma(3) + \frac{2\delta_T}{4\pi d}\sum_{m=1}^{\infty}\frac{1}{m^4}\exp\left(-\frac{4\pi md}{\delta_T}\right)\right). \quad (42)$$

Replacing the exponent with one, i.e. increasing the sum, we get approximately

$$\tilde{F} \approx \frac{k_B T}{4\pi d^3}\left(\varsigma(3) + \frac{2\delta_T}{4\pi d}\varsigma(4)\right),$$

that is, the result is two times less than Casimir's. Apparently, the two terms dropped in (40) provide a significant contribution. The value $\varsigma(3)$ here makes a small contribution compared to the sum, and it can be omitted, and formula (40) is an asymptotic expansion. In this case, the sum in (2) at $d/\delta_T \to 0$ turns, as shown above, into an integral, and the result according to [3] should transform into (1). The double sum in (40) diverges and should have an infinite limit, increasing as $\pi^3\delta_T/(60d)$ . Denoting $4\pi mnd/\delta_T = z_m$, we see that the points are located infinitely close, and the sum over m can be approximately replaced by an integral, $dm = \delta_T dz/(4\pi nd)$, $m = \delta_T z/(4\pi nd)$:

$$\sum_{m=1}^{\infty} = (4\pi n)^2 \frac{d^2}{\delta_T^2} \int_{\frac{4\pi nd}{\delta_T}}^{\infty} \exp(-z)\left(\frac{1}{z}+\frac{2}{z^2}+\frac{2}{z^3}\right)dz .$$

This result can be expressed in terms of the integral exponential and its primitives. However, the constructive approach is to integrate in parts, which gives the exact value of the integral. Integrate the first term:

$$\int_{\frac{4\pi nd}{\delta_T}}^{\infty} \exp(-z)\left(\frac{1}{z}+\frac{2}{z^2}+\frac{2}{z^3}\right)dz =$$

$$= \frac{\delta_T \exp(-4\pi nd/\delta_T)}{4\pi nd} + \int_{\frac{4\pi nd}{\delta_T}}^{\infty} \exp(-z)\left(\frac{1}{z^2}+\frac{2}{z^3}\right)dz .$$

Now, when re-integrating, the first term in the bracket destroys the second, and the integral is fully calculated. We have

$$\tilde{F}(d,T) = \frac{\hbar c}{4\pi d^3\delta_T}\left[\varsigma(3) + \sum_{n=1}^{\infty}\exp\left(-\frac{4\pi nd}{\delta_T}\right)\left(\frac{4\pi nd}{\delta_T}+1\right)\right].$$

In this result, the variable *n* also runs through an almost continuous range of values, therefore, using $dn = \delta_T dz/(4\pi d)$ and converting the sum into an integral, we obtain

$$\tilde{F}(d,T) \approx \frac{\hbar c}{4\pi d^3\delta_T}\left(\varsigma(3) + 2 + \frac{\delta_T}{2\pi d}\exp\left(-\frac{4\pi d}{\delta_T}\right)\right) . \qquad (43)$$

This result tends to $\hbar c/(8\pi^2 d^4)$, which is about three times less than Casimir's result. Apparently, this is due to the replacement of sums with singular integrals for small *z*, as well as the fact that the contour integral is not taken into account in (2). Indeed, if we change the order of conversion of sums into integrals, we get the result

$$\tilde{F}(d,T) \approx \frac{\hbar c}{4\pi d^3\delta_T}\Big[\varsigma(3) + \int_{4\pi d/\delta_T}^{\infty}\exp(-z)\Big(1 + 2z^{-1} + z^{-2}(2 + 8\pi d/\delta_T) + $$
$$+ z^{-3}8\pi d/\delta_T + 32 z^{-4}(\pi d/\delta_T)^2\Big)dz\Big] .$$

The integral in the limit diverges, but the result does not coincide with the previous one. Calculations using formula (40) are shown in Fig. 1. For small *d*, the number of terms in the sums increased proportionally to $1/d$. It can be seen that for small $d$, the result is always lower than the Casimir line. At long distances $d$, it always goes higher and tends to the line $k_B T\varsigma(3)/(4\pi d^3)$. Note that in [3], the result for the metal in formula (5.6) is two times less. This is a consequence of the loss of the first term in formula (5.3) of this work at $n=0$, when DP at low frequencies the value $\varepsilon_0$ was considered finite. This is true for dielectrics, but then a limit is applied for metals $\varepsilon_0 \to \infty$. The result (40) can be written in a more convenient form $\tilde{F}(d,T) = 2\hbar c S(d,\delta_T)/\delta_T^4$, where $S(d,\delta_T) = a_1\delta_T / d + a_2\delta_T^2 / d^2 + a_2\delta_T^3 / d^3$ and the coefficients are indicated:

$$a_1(d/\delta_T) = 2\pi\sum_{n=1}^{\infty}\sum_{m=1}^{\infty}\frac{n^2}{m}\exp\left(-\frac{4\pi mnd}{\delta_T}\right),$$

$$a_2(d/\delta_T) = \sum_{n=1}^{\infty}\sum_{m=1}^{\infty}\frac{n}{m^2}\exp\left(-\frac{4\pi mnd}{\delta_T}\right),$$

$$a_3(d/\delta_T) = \frac{\varsigma(3)}{8\pi} + \frac{1}{4\pi}\sum_{n=1}^{\infty}\sum_{m=1}^{\infty}\frac{1}{m^3}\exp\left(-\frac{4\pi mnd}{\delta_T}\right).$$

Thus, formula (2) does not correspond to the results of [18]. Namely, (38) from the formula (43) obtained from (2) by a multiplier, i.e., it is generally cardinal. The result of [18] gives the correct limit with $P_c$, and (43) is in 0.308 times smaller. This can be explained by the fact that (2) does not take into account the contour integral. The need to take into account the contour integral is discussed in [11].

## 4. Consideration of the contour integral

Consider formula (8) for $k_p \to \infty$, $r_{e,h}^2 = 1$ in the form

$$\tilde{F}'(d) = \frac{\hbar c}{\pi^2}\left(\frac{2}{\delta_T}\right)^4 p.v.\int_0^{\infty} z^3\cot(z)\sum_{m=1}^{\infty}\int_1^{\infty} p^2\exp(-a_m zp)dpdz, \tag{44}$$

Where $a_m = 4md/\delta_T$. The function $z^3\cot(z)\exp(-a_m zp)$ has poles at the points $z_l = l\pi$, $l = 1,2,...$. In their neighborhood, it can be approximated as $z_l^3\exp(-a_m z_l p)/(z - l\pi)$. The integral of this function over the domain $(l\pi - \pi/2, l\pi + \pi/2)$ in the sense of the main value is zero. The cotangent is a periodic function, so the integral can be transformed as a sum of integrals over the domain $(0,\pi)$:

$$\tilde{F}'(d) = \frac{\hbar c}{\pi^2}\left(\frac{2}{\delta_T}\right)^4 p.v.\int_0^{\pi} z^3\cot(z)\sum_{n=0}^{\infty}\sum_{m=1}^{\infty}\int_1^{\infty} p^2\exp(-a_m pz)\exp(-b_{mn} p)dpdz. \tag{45}$$

Here $b_{mn} = 4\pi mnd / \delta_T$. The cotangent in this area can be very accurately approximated by a function $\cot(z) = 1/z + 1/(z-\pi)$. By subtracting a function from the function $z^3 \cot(z)\exp(-a_m z)$ the function $\pi^3 yxp(-a_m \pi p)/(z-\pi)$, we remove a singularity. It remains to integrate into (42) the function $z^2 \exp(-a_m pz)$. Its integral is the following function *p*:

$$g_m(p) = \frac{2}{a_m^3 p^3}\left(1 - \exp(-a_m \pi p)\right) - \exp(-a_m \pi p)\left(\frac{\pi^2}{a_m p} + \frac{2\pi}{a_m^2 p^2}\right).$$

Its integral with function $p^2 \exp(-b_{mn} p)$ with respect to *p* for large $a_m$ and $b_{mn}$ can be taken in parts. Taking the approximation $g_m(p) \approx 2/(a_m^3 p^3)$ for small $a_m$ ($d << \delta_T$), we see that for $n$=0 the main contribution $2\ln(\delta_T / d)/(a_m^3)$ is obtained, and for the remaining *n* the contribution is $2\exp(-b_{mn})/(a_m^3 b_{mn})$ proportional to $\exp(-b_{mn})\delta_T^4 / d^4$, i.e. for the correction we have

$$\tilde{F}'(d) \approx \frac{4\hbar c}{\pi^5 \delta_T d^3}\left(\ln\left(\frac{\delta_T}{d}\right)\varsigma(3) + \frac{\delta_T}{32\pi d}\sum_{m=1}^{\infty}\sum_{n=1}^{\infty}\frac{\exp(-b_{mn})}{m^4 n}\right). \tag{46}$$

With an infinite upper limit $p_{\max}$, we have a divergence, which is understandable in principle: at $d$=0, the force is infinite. We took it as $p_{\max} = \delta_T / d$. We see that the sum can be converted into an integral, which gives

$$\tilde{F}'(d) \approx \frac{4\hbar c}{\pi^5 \delta_T d^3}\left(\ln\left(\frac{\delta_T}{d}\right)\varsigma(3) + \frac{1}{8}\left(\frac{\delta_T}{4\pi d}\right)^3 \exp\left(-\frac{4\pi\delta_T}{d}\right)\right). \tag{47}$$

For large distances $d >> \delta_T$, we have $g_m(p) \approx -\pi^2 \exp(-a_m \pi p)/(a_m p)$, and after a single integration in parts, discarding the remainder at $n$=0 , we get $-\pi^2 \exp(-a_m \pi)\left(1/(a_m^2 \pi) + 1/(a_m^3 \pi^2)\right)$, and for $n$>0 , respectively $-\pi^2 \exp(-a_m \pi - b_{mn})\left[(a_m(a_m \pi + b_{mn}))^{-1} + (a_m(a_m \pi + b_{mn})^2)^{-1}\right]$, and the result

$$\tilde{F}'(d) \approx -\frac{2\hbar c}{\pi \delta_T^2 d^2}\sum_{m=1}^{\infty}\frac{\exp(-4\pi md / \delta_T)}{m^2}\left(1 + \frac{1}{2}\sum_{n=1}^{\infty}\exp(-4\pi mnd / \delta_T)\right). \tag{48}$$

In it, we omitted the small terms proportional to the third powers of the coefficients. The sum of *n* can also be omitted. This correction is small in accordance with the fact that (2) gives the correct result at high temperatures limit.

**5. Temperature correction to DM**

We use the DSM in the form $\varepsilon(k) = \varepsilon_L + k_p^2 / (k_{00}^2 + k_c k + k^2)$. DM is obtained when $k_{00}^2 = 0$. Since small frequencies are interesting, DP will be decomposed by a small parameter.

When $k > k_c$ in the field $\delta_p < d < \delta_c$ we have $\varepsilon(k) = \varepsilon_L + k_p^2\left(1 - k_c k / \left(k_{00}^2 + k^2\right)\right) / \left(k_{00}^2 + k^2\right)$. DP is limited for DSM. When $k < k_c$ in the region $d > \delta_c$ we have $\varepsilon(k) = \varepsilon_L + k_p^2\left(1 - k^2 / \left(k_{00}^2 + k_c k\right)\right) / \left(k_{00}^2 + k_c k\right)$. The DP for DM $\varepsilon(k) = \varepsilon_L - k_p^2 / k_c^2 + k_p^2 / \left(k_c k\right)$ is large. The difference between DSM and PM now is that the value $K_\varepsilon = \sqrt{k^2 \varepsilon(k) + \kappa^2}$ is always complex. Considering the frequency spectrum up to PF, it can be considered $\varepsilon_L$ as a constant. The value $k_c$, generally speaking, should depend on the frequency and decrease with its increase. Indeed, for a short period of time, the electrons cannot deviate much from their average position. This ensures the transition to PM at a high frequency. It also depends on the temperature. You can take a dependency $k_c(T) = k_c(T_0) T / T_0$, which ensures a continuous transition to PM. If $k_0 < k_c$, then $K'_\varepsilon \approx p + k_0^2(\varepsilon_L - 1)/(2p)$, $K''_\varepsilon \approx -k_p^2 k_0 / (2pk_c)$. These values can have different orders of magnitude for small and large *p*. If $k_0 > k_c$, then $K'_\varepsilon = \sqrt{p^2 + k_0^2(\varepsilon_L - 1) - k_p^2 / k_0^2}$, $K''_\varepsilon = -k_p^2 k_c / \left(2k_0^3 K'\right)$, if $p^2 > k_p^2 / k_0^2 - k_0^2(\varepsilon_L - 1)$. Otherwise $K'_\varepsilon = -k_p^2 k_c / \left(2k_0^3 K''\right)$, $K''_\varepsilon = -\sqrt{k_p^2 / k_0^2 - k_0^2(\varepsilon_L - 1) - p^2}$. This leads to very complex decompositions that require taking into account all integrals, and for different ranges of *p* values. The absence of poles in $\varepsilon(0)$, i.e. a large DP does not allow the use of decompositions. It is simpler and more accurate to use the calculation of the integral over *p* in (1) and (2) by integrating in parts. We use (2) as

$$P_\mu(d,T) = -\frac{k_B T}{\pi} \sum_{n=0}^{\infty} \sum_{m=1}^{\infty} \int_1^{\infty} \frac{k_n^3}{1 + \delta_{n0}} r_{\mu n}^{2m}(p) \exp(-2mpk_n d) p^2 dp.$$

The transition to (1) is carried out by replacing the sum with $k_n$ integration with *k*. The integral is calculated as

$$I_{\mu n}(m,d) = \int_1^{\infty} \frac{k_n^3}{1 + \delta_{n0}} r_{\mu n}^{2m}(p) \exp(-2pmk_n d) p^2 dp \approx \exp(-2mk_n d) \times$$
$$\times \left[ \frac{k_n^2 r_{\mu n}^{2m}(1)}{(2k_n md)} + k_n \frac{2r_{\mu n}^{2m}(1) + \partial_p r_{\mu n}^{2m}(1)}{(2md)^2} + \frac{2r_{\mu n}^{2m}(1) + 4\partial_p r_{\mu n}^{2m}(1) + \partial_p^2 r_{\mu n}^{2m}(1)}{(2md)^3} \right], \quad (49)$$

that is, we discard the residual integral, which we consider to be small. The indicated integral is small in the far zone because it is proportional to $1/d^4$. For even powers of RC $r_{\mu n}^{2m}(1) = r_\mu^{2m}(k_n, 1)$, and their derivatives $\partial_p r_{\mu n}^{2m}(1) = \partial_p r_\mu^{2m}(k_n, 1)$, we have Appendix formulas (A9)–(A16). In particular,

$$r_\mu^{2m}(k,1)=\left(\frac{1-\varepsilon^{-1/2}(k)}{1+\varepsilon^{-1/2}(k)}\right)^{2m} .$$

$$\partial_p r_e^{2m}(k,1)=\frac{4m}{\varepsilon^{1/2}(k)}\frac{\left(1-\varepsilon^{-1/2}(k)\right)^{2m}}{\left(1+\varepsilon^{-1/2}(k)\right)^{2m}} ,$$

$$\partial_p r_h^{2m}(k,1)=-\frac{4m}{\varepsilon^{1/2}(k)}\frac{\left(1-\varepsilon^{1/2}(k)\right)^{2m-1}}{\left(1+\varepsilon^{1/2}(k)\right)^{2m+1}} ,$$

which we take when $k=k_n$. We have $r_\mu^{2m}(0,1)=1$, $\partial_p r_\mu^{2m}(0,1)=0$, $\partial_p^2 r_\mu^{2m}(0,1)=0$. This results $1/(2md)^3$ in the following square bracket value at (49) at $n$=0. Summation over m gives $\varsigma(3)/(2d)^3$. It is important that $r_{e0}^{2m}(1)=r_{h0}^{2m}(1)=1$, $\partial_p r_{\mu n}^{2m}(1)=0$, $\partial_p^2 r_{\mu n}^{2m}(1)=0$. For DM, there is a simple pole at zero for DP, for PM there is a second–order pole, and for DSM there is no pole. In the discarded integral at $n$=0, we have terms $\partial_p^4\left(p^2 r_{\mu 0}^{2m}\right)=2\partial_p^2 r_{\mu 0}^{2m}+...$ that have stronger attenuation at large $d$. It is the order that we discard. The same applies to terms with n>0. Selecting a term with $n$=0, we have the result

$$P_{e,h}(d,T)=-\frac{k_B T}{\pi}\left(\frac{\varsigma(3)}{8d^3}+\sum_{n=1}^{\infty}\sum_{n=0}^{\infty}k_n^3 I_{(e,h)n}(m,d)\right). \tag{50}$$

It is suitable for all DP models, but the simplest result is obtained for PM. For DM, it will have additional dependencies on $d$ in parenthesis. We see that for all DP models there are dependencies on the inverse degrees of $d$. This means their justice in the far zone. For the ratio (50), this is understandable, since it is obtained by integration in parts, i.e. expansion in inverse powers of $d$. For formula (1), the result can be obtained for any finite distance $d$, however, with its decrease, the limits of integration should increase, ensuring the convergence of the integral. At $d$=0, the integral diverges logarithmically. Formally, additional convergence can be provided by the DP model with its faster tendency to unity than according to the law $1+k_p^2/k^2$. However, this is a non-physical result. DM gives the pole of DP of the first order in k, and PM of the second order. PM has the fastest tendency to infinity. For a finite $p$, we have $r_e^2(k,p)=\left[k_p^4/\left(4k^4\right)\right]\left(1/\left(2p^2\right)-1\right)^2$. However, at $d$=0, there is a logarithmic divergence, which indicates that formula (1) is inapplicable in this case. And have a similar relationship for $r_h^2(k,p)$ .

## 6. About the van Kampen-Schram formula

Formula (1) can be written in the van Kampen-Schram form [10–14]

$$P_\mu = -\frac{\hbar c}{2\pi^2}\int_0^\infty\int_0^\infty \frac{\tilde{k}_z \kappa d\kappa dk}{f_\mu(k,\kappa)} = -\frac{\hbar c}{2\pi^2}\int_0^\infty k^3 \int_1^\infty \frac{p^2 dp dk}{f_\mu(k,p)} = \\ = -\frac{\hbar c}{32\pi^2 d^4}\int_1^\infty \frac{dp}{p^2}\int_0^\infty \frac{y^3 dy}{r_\mu^{-2}(y,p)\exp(y)-1} . \quad (51)$$

$f_\mu(k,\kappa)$=0 are the dispersion equations for PP for an arbitrary plane-layered structure. In the case of a gap with an extended flat-layered structure $f_\mu(k,\kappa) = r_\mu^{-2}(k,\kappa)\exp(2\tilde{k}_z d) - 1$. In the case of a limited volume they correspond to poles that determine its resonant frequencies. Casimir in [17] considered a model problem for a limited resonator with ideal walls and dimensions $L_x = L_y = L$, $L_z = a$. In such a resonator, only radiated modes forming resonances $\omega_n = c\sqrt{(k\pi/L)^2 + (kl/L)^2 + (m\pi/a)^2}$ with a multiindex $n = (klm)$ are possible. In a resonator with dissipative walls, the frequencies become complex. Casimir summed up the energy of zero oscillations $\hbar\omega_n/2$, which naturally turned out to be infinite. Next, he moved to the limit $L \to \infty$, i.e., to an open Fabry-Perrault resonator with ideal walls (screens). PP is not supported in such a resonator. However, with impedance screens, slow surface PP and fast radiated modes appear in it. The former are responsible for the force in the near zone, while the latter provide the main contribution in the far zone [14]. As a result of Casimir, there were no surface PP, so it is typical for the far zone. Indeed, to determine the additional energy resulting from the boundary conditions, he subtracted the energy from the sum at an infinitely large gap ( ). It also turned out to be endless. This means that with infinitesimal d=a and infinite conductivity, there are no zero fluctuations in the gap, and outside it the energy is infinite. Infinitely conductive plates would contract with infinite force. However, there are no such materials in nature. To overcome this, Casimir assumed that for real metals, RC values are close to minus one at significantly low frequencies, and in the region of some high frequency such as plasma, they abruptly go to zero. In the final formula for $\delta E$, which is then summed using the Euler-Maclaurin formula, a large value of $a$ is tacitly assumed. Thus, the Casimir formula is correct for metals in the far zone $k_p d >> 1$ when the condition for RC squares is met with a good degree of accuracy $r_{e,h}^2 \approx 1$ at low frequencies. Their deviation from unity allows you to determine the corrections. In the case where resonant frequencies are determined from $f_\mu(k,\kappa) = 0$, formula (51) means summing the energy of zero oscillations according to the “principle of argument” theorem: the integral along the contour of the imaginary axis – the right semicircle gives the specified amount, given that the Casimir free energy is determined through the integral of $\ln(f_\mu(k,\kappa))$ [12]. In this case, the imaginary parts of the frequencies are complexly conjugated and annulled. In the case of an open resonator, the discrete frequencies are transformed into continuous dispersion relations, their imaginary parts are annulled, and the sum is reduced to the integral (51). Formula (51) is remarkable because it allows us to consider arbitrary multilayer plane-layered structures with any number of layers, i.e. generalizes (1). In Section 2, we assumed $r_{e.h}^2(k,p)=1-\alpha_{e,h}(k,p)$ where

$$\alpha_{e,h}(k,p)=\left(1-\frac{2\beta_{e,h}}{1+\beta_{e,h}}\right)^{2m}, \tag{52}$$

$$\beta_e=\frac{\sqrt{\varepsilon(k)-1+p^2}}{p\varepsilon(k)}, \tag{53}$$

$$\beta_h=\frac{p}{\sqrt{\varepsilon(k)-1+p^2}}. \tag{54}$$

Here we will once again consider obtaining corrections by decomposing into small parameters based on the smallness of the quantities $\beta_{e,h}$. Decomposition by small parameters allows the analytical calculation of integrals. Otherwise, rather complex integration in parts should be used. For the smallness $\beta_e$ condition is $p>1/\sqrt{\varepsilon(k)+1}$, which is always fulfilled. The value $\beta_h$ is always less than one, but tends to it at high frequencies. EFor its smallness, it is necessary to require $k\le k_p$, $p\le p_{\max}=k_p/k$ (for PM). For DM $p_{\max}=k_p/\sqrt{k^2+kk_c}$ and for DSM $p_{\max}=k_p/\sqrt{k_{00}^2+k^2+kk_c}$, where $k_{00}^2=\omega_{00}^2/c^2$. Generally speaking, these upper limits should be set in integrals when calculating the contribution of magnetic modes. This leads to cumbersome formulas and additional terms. However, we will not do this, justifying this by the fact that taking into account infinite limits leads to contributions with large negative powers of *d*, which are significantly less than for electric modes. Since the integral over *p* converges at an infinite upper limit, the latter is convenient to use. At extremely high *p* for electric modes $\beta_e\approx 1/\varepsilon(k)$, i.e. at low frequencies, this value is small, but for magnetic modes it is on the order of unity $\beta_h\approx 1$. This suggests that magnetic modes do not contribute in this case. Now for the DSM we have decompositions

$$\beta_e=\frac{\sqrt{k_{00}^2+k^2+kk_c}}{pk_p}\left(1+\frac{p^2\left(k_{00}^2+k^2+kk_c\right)}{2k_p^2}\right), \tag{55}$$

$$\beta_h=\frac{p\sqrt{k_{00}^2+k^2+kk_c}}{k_p}\left(1-\frac{p^2\left(k_{00}^2+k^2+kk_c\right)}{2k_p^2}\right). \tag{56}$$

For DM these formulas the $k_{00}^2$ should be omitted, and for PM the $k_c = 0$ should also be taken. Formulas (55), (56) can be written in general form $\beta_e = \beta_{e0}(k)/p + \beta_{e1}(k)p$, $\beta_h = \beta_{h0}(k)p - \beta_{h1}(k)p^3$. Considering the values (30) and (31) small, from (27) we have

$$r_{e,h}^{2m}(k,p) = \left(1 - \frac{2\beta_{e,h}}{1+\beta_{e,h}}\right)^{2m} \approx \left(1 - 2\beta_{e,h} + 2\beta_{e,h}^2\right)^{2m} \approx$$
$$\approx 1 - 4m\beta_{e,h}\left(1 - \alpha_{e,h}\right) + 4m(2m-1)\left(\beta_{e,h} - \beta_{e,h}^2\right)^2 - \ldots.$$

Taking the terms up to the second order, we get

$$r_{e,h}^{2m}(k,p) \approx 1 - 4m\beta_{e,h} + 8m^2\beta_{e,h}^2 \ . \qquad (57)$$

It remains to calculate the integrals (51). If $r_{e,h}^2 = 1$, this results in the Casimir result that does not depend on the order of integration. First, we will integrate over *p*, and then over *k*. When expanding (51), integrals arise

$$I_{\mu m}(k,d) = \int_1^\infty r_\mu^{2m} \exp(-2mpkd)p^2 dp.$$

The simplest result is obtained by using the first order:

$$I_{em}(k,d) = \exp(-2mkd)\times$$
$$\times\left[\frac{1}{2mkd} + \frac{2}{(2mkd)^2} + \frac{2}{(2mkd)^3} - 4m\beta_{e0}(k)\left(\frac{1}{2mkd} + \frac{1}{(2mkd)^2}\right) - \right.$$
$$\left. - 4m\beta_{e1}(k)\left(\frac{1}{2mkd} + \frac{3}{(2mkd)^2} + \frac{6}{(2mkd)^3} + \frac{6}{(2mkd)^4}\right)\right], \qquad (58)$$

$$I_{hm}(k,d) = \exp(-2mkd)\times\left\{\frac{1}{2mkd} + \frac{2}{(2mkd)^2} + \frac{2}{(2mkd)^3} - \right.$$
$$-4m(\alpha_{h0}(k))\left[\frac{1}{2mkd} + \frac{3}{(2mkd)^2} + \frac{6}{(2mkd)^3} + \frac{6}{(2mkd)^4}\right] +$$
$$\left. + 4m\alpha_{h1}(k)\left[\frac{1}{2mkd} + \frac{5}{(2mkd)^2} + \frac{20}{(2mkd)^3} + \frac{60}{(2mkd)^4} + \frac{120}{(2mkd)^5} + \frac{120}{(2mkd)^6}\right]\right\}. \qquad (59)$$

After summation and integration over *k*, the terms independent on $\beta_{e,h}$ give a Casimir result with dependence $1/d^4$. For PM $\beta_{e0} = \beta_{h0} = k/k_p$, $\beta_{e1} = \beta_{h1} = k^3/2k_p^3$, therefore, integer powers of *k* arise in the corrections, which leads, after integration over k, to dependencies $1/d^n$, n=4.5, ... For DM and DSM $\beta_{e0} = \beta_{h0} = \sqrt{k_{00}^2 + k^2 + kk_c}/k_p$, $\beta_{e1} = \beta_{h1} = \left(k_{00}^2 + k^2 + kk_c\right)^{3/2}/\left(2k_p^3\right)$, which leads to integrals for which tabular values are expressed in terms of special functions.

Considering the frequency to be small, for DM we take $\beta_{e0}=\beta_{h0}\approx\sqrt{kk_c}\left(1+k/2k_c\right)/k_p$, $\beta_{e1}=\beta_{h1}\approx\left(kk_c\right)^{3/2}\left(1+2k/\left(2k_c\right)\right)/\left(2k_p^3\right)$, which leads to fractional powers $1/d^{n+1/2}$ in correction units. For the DSM $\beta_{e0}=\beta_{h0}\approx k_{00}\left(1+3\left(k^2+kk_c\right)/\left(2k_{00}^2\right)\right)/k_p$, $\beta_{e1}=\beta_{h1}=k_{00}^3\left(1+3\left(k^2+kk_c\right)/\left(2k_{00}^2\right)\right)/\left(2k_p^3\right)$, which leads to a decrease in the contribution from the term $1/d^4$ and to integer powers $1/d^n$, $n\geq5$. The relations allow us to obtain an exact correction term with $n$=5 and a contribution to the term with $n$=7. To obtain term with dependency $1/d^6$, the second order in (57) must be taken into account. It is given by additional terms, formulas (A17), (A18) in Appendix. The expansion for the force is obtained by integrating and summing the quantities $k^3I_{\mu m}\left(k,d\right)$ and $k^3\delta I_{\mu m}\left(k,d\right)$.. The results for PM with $\varepsilon_L=1$ are given in the Appendix, formulas (A19)–(A22). For PM, the first two coefficients are obtained as $c_1=16/3, a_2=24$, which are established in many works. To do this for $\varepsilon_L\neq1$, use (28) and (29). The formulas make it possible to obtain decompositions for any DP function, as well as higher-order decompositions and contributions with larger negative powers. Note that the formulas are valid for $k_pd>>1$, therefore, the meaning of obtaining higher terms is small. For DM, taking into account one member, we have $P(d)=P_C\left(1-c_1\left(k_pd\right)^{-1/2}\right)$, and

$$c_{1/2}=-2^{9/2}\frac{3\Gamma\left(4+1/2\right)\varsigma\left(3+1/2\right)}{5\Gamma\left(4\right)\varsigma\left(4\right)}.$$

The formula is based on decompositions of the form

$$\beta_e=\frac{\left(k_ck\right)^{1/2}+k_c^{-1/2}k^{3/2}/2}{pk_p}+p\frac{\left(k_c^{1/2}k^{5/2}/2+\left(k_ck\right)^{3/2}+k_c^{-1/2}k^{7/2}/2\right)}{2k_p^3},$$

$$\beta_h=p\frac{\left(k_ck\right)^{1/2}+k_c^{-1/2}k^{3/2}/2}{k_p}-p^3\frac{\left(k_c^{1/2}k^{5/2}/2+\left(k_ck\right)^{3/2}+k_c^{-1/2}k^{7/2}/2\right)}{2k_p^3},$$

and it is not applicable for small $k_c$.

## 7. Integration in parts at T=0

Using decompositions by small parameters are used above, we obtain the decomposition (51) by integrating it by parts. We have the result (49), in which we should replace $k_n\to k$ and $I_{\mu n}\left(m,d\right)\to I_\mu\left(m,k,d\right)\exp\left(-2mkd\right)$, $r_{\mu n}^{2m}\left(p\right)\to r_\mu^{2m}\left(k,p\right)$. Thus,

$$P_\mu\left(d\right)=-\frac{\hbar c}{2\pi^2}\sum_{m=1}^{\infty}\int_0^\infty k^3I_\mu\left(m,k,d\right)\exp\left(-2mkd\right)dk. \qquad (60)$$

This is a decomposition in inverse powers of $d$ with coefficients depending on $d$, where

$$k^3 I_\mu(m,k,d) = \frac{r_\mu^{2m}(k,1)k^2}{(2md)} + \frac{2r_\mu^{2m}(k,1) + \partial_p r_\mu^{2m}(k,1)}{(2md)^2}k + \\ + \frac{2r_\mu^{2m}(k,1) + 4\partial_p r_\mu^{2m}(k,1) + \partial_p^2 r_\mu^{2m}(k,1)}{(2md)^3} . \quad (61)$$

Integral (60) can be calculated by decomposing RC and their derivatives in $k$ or numerically. Since $r_{e,h}^{2m}(0,1) = 1$, $r_{e,h}^{2m}(\infty,1) = 0$, $\partial_p r_{e,h}^{2m}(0,1) = 0$, $\partial_p^2 r_{e,h}^{2m}(0,1) = 0$, it can also be calculated by integration in parts over the $k$ variable. Integrating the first term in (61) four times in parts, we have

$$\int_0^\infty \frac{r_\mu^{2m}(k,1)k^2}{(2md)} \exp(-2mkd)dk = \frac{2r_\mu^{2m}(k,1)_{k=0}}{(2md)^4} + \\ + \frac{6\partial_k r_\mu^{2m}(k,1)_{k=0}}{(2md)^5} + \int_0^\infty \frac{\partial_k^4 \left(k^2 r_\mu^{2m}(k,1)\right)}{(2md)^5} \exp(-2mkd)dk \approx . \\ \approx \frac{2}{(2md)^4} + \frac{6\partial_k r_\mu^{2m}(k,1)_{k=0}}{(2md)^5}$$

We omitted the remaining integral. Integrating the second term three times, omitting the remaining integral and considering $\partial_p r_\mu^{2m}(k,1)_{k=0} = 0$, we obtain

$$\int_0^\infty \frac{2r_\mu^{2m}(k,1) + \partial_p r_{\mu n}^{2m}(1)}{(2md)^2} \exp(-2mkd)kdk \approx \\ \approx \frac{2}{(2md)^4} + \frac{4\partial_k r_\mu^{2m}(k,1) + 2\partial_k \partial_p r_\mu^{2m}(k,1)}{(2md)^5} |_{k=0} .$$

Integrating the third term twice, we get

$$\int_0^\infty \frac{2r_\mu^{2m}(k,1) + 4\partial_p r_{\mu n}^{2m}(k,1) + \partial_p^2 r_\mu^{2m}(k,1)}{(2d)^3} \exp(-2mkd) \approx \\ \approx \frac{2}{(2md)^4} + \partial_k \frac{2r_\mu^{2m}(k,1) + 4\partial_p r_{\mu n}^{2m}(k,1) + \partial_p^2 r_\mu^{2m}(k,1)}{(2md)^5} |_{k=0}$$

We omitted the remaining integrals, which contribute to terms with higher negative powers of $d$ and which we consider small for large $d$. However, to get the correct result, one more integral should be taken into account in the first order. We calculate the derivatives according to the formulas of the Application:

$$\partial_k r_\mu^{2m}(k,1) = 2m \frac{\left(1 - \varepsilon^{-1/2}(k)\right)^{2m-1}}{\left(1 + \varepsilon^{-1/2}(k)\right)^{2m+1}} \frac{\partial_k \varepsilon(k)}{\varepsilon^{3/2}(k)}, \quad (62)$$

$$\partial_k \partial_p r_e^{2m}(k,1) = 2m \frac{\left(1 - \varepsilon^{-1/2}(k)\right)^{2m}}{\left(1 + \varepsilon^{-1/2}(k)\right)^{2m}} \left[(4m+1)\varepsilon^{-1/2}(k) - 1\right] \frac{\partial_k \varepsilon(k)}{\varepsilon^{3/2}(k)}, \quad (63)$$

$$\partial_k \partial_p r_h^{2m}(k,1) = -\partial_k \partial_p r_e^{2m}(k,1). \tag{64}$$

There is $\partial_k r_\mu^{2m}(0,1) = -4m/k_p$ for PM and for DM the result $\partial_k r_\mu^{2m}(0,1) = -2mk_c^{3/2}/\left(k_p k^{1/2}\right)_{k=0} \to \infty$ is divergent. Thus, DM is not correct for this method. The solution here is to use the DSM, for which all values are finite. Another way out is to use small parameter expansions in (61) with direct integration in (60) instead of integration by parts. For low frequencies $r_\mu^{2m}(k,1) \approx 1 - 4m\varepsilon^{-1/2}(k)$, therefore, differentiating this formula gives the same result $\partial_k r_\alpha^{2m}(0,1) \approx 2m\varepsilon^{-3/2}(k)\partial_k \varepsilon(k)_{k\to 0}$ as differentiating the exact formula. This also applies to the differentiation of derivatives: they all have a singularity $\varepsilon^{-1/2}(k)$, when differentiating for PM gives $\partial_k \partial_p r_e^{2m}(0,1) = -\partial_k \partial_p r_h^{2m}(0,1) = -2m\varepsilon^{-3/2}(0)\partial_k \varepsilon(0) = 4m/k_p$. Differentiation of the second derivatives gives for the electric mode $\partial_k \partial_p^2 r_e^{2m}(0,1) = 4m\varepsilon^{-3/2}(k)\partial_k \varepsilon(k) = -8m/k_p$ and $\partial_k \partial_p^2 r_h^{2m}(0,1) = -16m^2\varepsilon^{-2}(k)\partial_k \varepsilon(k) = 0$ for the magnetic mode, since for it the singularity of the second derivative is stronger $\varepsilon^{-1}(k)$. From this it can be seen that the reason for the divergence in DM is the slow increase of DP at zero. For the DSM $\partial_k r_\mu^{2m}(0,1) = -2mk_c/\left(k_p k_{00}\right)$, and there is no divergence, but it does occur during the transition to DM when $k_{00} \to 0$. When differentiating derivatives in the DSM, multipliers of the type $\left(1 - k_{00}^2/k_p^2\right)^{2m}/\left(1 + k_{00}^2/k_p^2\right)^{2m}$ appear which very close to unity. For PM, considering that $r_\mu^{2m}(0,1) = 1$, $\partial_p r_\mu^{2m}(0,1) = 0$, $\partial_p^2 r_\mu^{2m}(0,1) = 0$, and the ratios for the derivatives, we get a correction factor 11/3 instead of 16/3. This is explained by the fact that (49) contains an insufficient number of expansion terms. Exactly, the following terms $\left[k^3 \partial_p^3\left(p^2 r_\mu^{2m}\right)/(2mkd)^4\right]_{k=0}$ should be considered. They do not contribute to the $1/d^4$ dependence (only the first three terms in the expansion (61) contribute to it), but together with the two terms in (61) they contribute to the $1/d^5$ dependence, thus obtaining the coefficient 16/3 and the contribution to the $1/d^6$ dependence. In order to accurately obtain the coefficient for this dependence, other terms of the decomposition should be taken into account. In this case, rather cumbersome expressions for derivatives $\partial_p^l\left(p^2 r_\mu^{2m}\right)_{p=1}$, $l \geq 3$ arise. The integrals omitted in (61) have the form

$$\int_1^\infty \frac{\partial_p^3\left(p^2 r_\mu^{2m}\right)}{(2md)^3} \exp(-2mkpd)\,dp,$$

where for derivatives we take the decomposition of RC for PM:

$$\partial_p^3\left(p^2 r_e^{2m}(k,p)\right)\approx -12m\frac{k_p k^3}{k_p^2\left(k^2+k_p^2\right)}+48m^2\frac{k^6 p}{k_p^2\left(k^2+k_p^2\right)^2}\approx ,$$
$$\approx 12mk^3/k_p^3+48m^2k^6p/k_p^6$$
$$\partial_p^3\left(p^2 r_h^{2m}(k,p)\right)\approx -mk\left(24-120k^2p^3/k_p^2\right)/k_p+ +192m^2pk^4/k_p^4 .$$

The contribution with the minimum degree $d$ in the denominator is given by the magnetic mode. After integration with respect to $k$, it is equal to $-24/\left[(2md)^4 k_p d\right]$, which just provides a coefficient of 16/3. For decomposition, we require $p\le p_{\max}(k)=k_p/k$. With a large PF, it is quite possible to use an infinite upper limit for $p$, which simplifies the results. The above result corresponds to PM. In the general case, the correction term depends on the DP model and can be written in the form $-\delta/d$ where the coefficient having the length dimension is

$$\delta=\frac{-1}{12\varsigma(4)}\sum_{\substack{m=1\\ \mu=e,h}}^{\infty}\partial_k\left(\frac{10r_\mu^{2m}(k,1)+\partial_p^2 r_\mu^{2m}(k,1)}{2m^5}\right. \left.+\frac{6\partial_p r_\mu^{2m}(k,1)+6\partial_p^2 r_\mu^{2m}(k,1)+\partial_p^3 r_\mu^{2m}(k,1)}{4m^6k}\right)_{k=0} .$$

Note that if, instead of integration by parts in, we take the expansion of RC in the vicinity of zero and substitute it in (60), we get a coefficient of 13/3 instead of 11/3 plus a contribution to the coefficient at $\left(k_p d\right)^{-2}$. The integration by part method allows calculating force corrections (51) with arbitrary dispersion relations for any DP model. It does not require explicit decompositions. They are obtained by sequential application of integrations. However, it involves very cumbersome calculations. The residual integral should also be evaluated.

## 8. Conclusion

The main results of the work are as follows. Formula (1) should be considered accurate, and formula (2) approximate, yielding results at high temperatures. At low temperatures, it does not go into (1). This is due to the fact that it does not take into account the contour integral. In section 4, the contour integral is calculated, but approximately based on the cotangent approximation. Corrections to the Casimir force due to DP of the real metal and the final temperature strongly depend on the DP model. In reality, the specific force is affected by the decreasing extremity of the interacting samples and the internal field, which significantly increases it. Based on (1), force corrections for PM and DM are obtained up to four orders of magnitude. In the far zone $d>\delta_c$, DM should be used, and in the region $\delta_p<d<\delta_c$, PM should be used taking into account the Lorentz term. He changes the coefficients in the expansion, starting with the third one. In the UV range, the Drude-Lorentz model should be used. We assume that the Drude term is important only for distances of the order $\lambda_p$, and for smaller ones it can be omitted. The results of the calculation using formula (1) are shown in Fig. 2. Based on formula (7), using the hyperbolic cotangent expansion, temperature corrections for ideal mirrors are obtained and the results are presented taking into account the first order for PM. Based on formula (8), the results of estimates of the contour integral for high and low temperatures are presented. The limits of applicability of the obtained ratios are established. The results obtained on the basis of formula (2) are shown in Fig. 1. Figure 3 shows numerical results for dielectric and metal plates of finite thickness, respectively. Metal plates are described by the Drude model. In Fig. 4 shows the results for the interaction of plates described by the Onsager-Drude-Smith model, PM and DM. The results can be obtained for any, even small, but finite *d*. At the same time, the smaller *d*, the higher the frequency upper limit should be taken when calculating an integral with an infinite upper limit. For any fixed *d*, it is always possible to estimate the discarded term and achieve its smallness, which was done in the calculations. The numerical curve for small *d* and *T*=0 is significantly lower than the Casimir result (for *d*=1 nm by two orders of magnitude), and for large *d* it tends to it Fig. 2. If there is an ideal dielectric with DP independent of frequency between the metal plates, the force decreases several times as $\sqrt{\tilde{\varepsilon}}$. This can be interpreted as an increase in the field energy in the gap, as a result of which the energy difference outside the plates and in the gap becomes smaller. Indeed, $k_z = \sqrt{k_0^2\tilde{\varepsilon} - \kappa^2}$ and possible substitutions $\kappa^2 = k^2\tilde{\varepsilon}\left(p^2 - 1\right)$, $\tilde{k}_z = k\sqrt{\tilde{\varepsilon}}\, p$, $\exp\left(2\tilde{k}_z d\right) = \exp\left(2kp\sqrt{\tilde{\varepsilon}}d\right)$, $k = y/\left(2p\sqrt{\tilde{\varepsilon}}\,d\right)$, which lead to this result.

The work [35] shows a significant effect of temperature on the force at not too large distances. This is demonstrated by experiments [35], as well as Fig. 1. Interestingly, all curves (including the Casimir result) are close in the region of *d*~1 μm. At short distances, they are significantly lower than the Casimir result, even at extremely low temperatures, which indicates the approximate nature of formula (2). At large distances, they go significantly higher than the Casimir result and have an asymptotic value of $1/d^3$. In the near zone, the temperature has almost no effect. Its influence is strong in the far zone. Curve 1 in Fig. 2 corresponds well to curve 1 in Fig. 1.

In the case of metal plates of finite thickness *t*, the van Kampen-Schram formula allows us to find the pressure depending on the thickness. At low thickness, it is proportional to *t*, but already at thicknesses of the order of tens of nm, it reaches saturation Fig. 3. The results for PM are given in [36,37]. Figure 3 shows the results for the Drude-Lorentz model for a dielectric and a metal. The result for the metal at $t$=100 nm, $d$=10 nm (curve 5) corresponds to the result of Fig. 2 at $d$=10 nm. The choice of DP model is essential. Most of the results in the literature are given for PM, for example, [15,18–23,36,37].

For real plates of finite transverse dimensions, when interacting at very large distances, it seems that the DSM is more preferable [31,32]. This is due to the fact that low-frequency currents vanish at the edges of the plate, so the specific force should depend on the transverse dimensions of the plates and decrease with decreasing $L$. If we take the temperature-dependent emergency, which vanishes at $T$=0, we can make a continuous transition to PM. In this case, the thermal theorem of Nernst is preserved. Taking into account the internal field can lead to a significant change in strength. For $d$=1 nm or less, the van der Waals theory of forces should be used. Figure 4 shows the temperature dependences of the force calculation for PM, DM, DSM and the Drude-Onsager model according to formula (2). The latter model increases the force density, while the DSM significantly reduces it for small sample sizes. When integrating over $p$, the upper limit can be taken as $p_{\max}(n) = k_p / k_n$, which greatly reduces the calculations. During the calculations, three hundred members of the series were taken and $p_{\max}(n) = 2 + 10k_p / k_n + 0.1/d$. It is always for DSM $r_\alpha^{-2} > 1$, so the term with $n$=0 drops out. For DM in this case $r_e^{-2} > 1 + 2(k_0 / k_p)^{1/2} / p$, $r_e^{-2} > 1 + 2p(k_0 / k_p)^{1/2}$, and for PM $r_e^{-2} > 1 + 2(k_0 / k_p)/ p$. It should be assumed here $k_0 = 0$, therefore , that the singularity that occurs at $n$=0 is excluded by the term $k_0^3$. The term with $n$=0 occurs only in the case of ideal conductivity. It is this that gives the dependence $1/d^3$. This dependence is approximately fulfilled for when the exponents are close to unity. For $T$=5 K, this is an area up to 0.46 mm, and for $T$=300 K, this is an area up to 8 μm. It is in this area that the force decreases more slowly than according to the law $1/d^4$ Fig. 4. However, at $d / \delta_T >> 1$ there is an influence of strongly attenuating exponentials in the series, so a sharp decrease in strength begins with distance. The lower the temperature, the further away the area of such a sharp attenuation appears, and at $T \to 0$ it goes to infinity, i.e. the $1/d^3$ dependence occurs in the far zone. This result, apparently, has not been noted in the literature. As for the differences between the models, they are invisible in Fig. 4. However, for $d$=5 nm and $d$=40 μm, the models give maximum and minimum values (in N/m2), respectively: 4512.8, $3.74 \cdot 10^{-25}$ (PM); 4450.2, $3.66 \cdot 10^{-25}$ (DM); 4513.4, $3.75 \cdot 10^{-25}$ (DSM), 9157.5, $3.97 \cdot 10^{-25}$ (Onsager-Drude-Smith model), i.e. a difference in the order of magnitude is possible twice. When calculating according to the DSM, $L$=10 μm was taken.

If we consider a separate metal plate of thickness $d$, then the field penetrates weakly into it, due to which an external compressive Casimir pressure arises. It can be explained as the van der Waals attraction between atoms. To obtain it, consider the RC from the side of the plate from its boundary with the vacuum. This changes the KO sign. But now $k_z = \sqrt{k_0^2 \varepsilon(\omega) - \kappa^2}$, the pressure is also changing significantly. It is not difficult to show that for large $d$ it decreases exponentially according to the power law, i.e. faster than $1/d^4$. This can be explained by the strong attenuation of the magnitude $\exp\left(-2k\varepsilon^{1/2}(k)d\right)$.

### Financing the work

The study was carried out with the financial support of the Ministry of Education and Science of the Russian Federation as part of a state assignment (project No. FSRR-2026-0006).

### Conflict of interest

The author reports that there is no conflict of interest.

### Application

For PM with $\varepsilon_L \neq 1$ we have

$$\begin{gathered} B(k,p) = \frac{k}{k_p}\left(p + \frac{1}{p}\right) + 4\frac{k^4}{k_p^4}\left(p^4 + \frac{1}{p^4}\right) + \\ + \frac{k^3}{2k_p^3}\left((2-\varepsilon_L)p - \frac{(\varepsilon_L+1)}{p} - 11p^3 - \frac{10}{p^3}\right) + \\ + \frac{k^5}{k_p^5}\left(\frac{p^3}{8}(1 + 66(\varepsilon_L - 1)) - \frac{2\varepsilon_L - 1}{4}p - \frac{\varepsilon_L(\varepsilon_L-1)}{2p} + \right. \\ \left. + 3\frac{p^8 + p^4(\varepsilon_L-1)^2}{8} - \frac{15}{2}\left(1 - \frac{(\varepsilon_L+1)}{p^2} - p^5\right)\right), \end{gathered} \tag{A1}$$

$$\begin{gathered} C = \frac{k^2}{k_p^2}\left(p^2 + \frac{1}{p^2}\right) - 4\frac{k^3}{k_p^3}\left(p^3 + \frac{1}{p^3}\right) + \\ + \frac{k^4}{k_p^4}\left(1 - \frac{(\varepsilon_L+1)}{p^2} + 3p^4 + \frac{4}{p^4} - p^2(\varepsilon_L - 1)\right) - \\ - 6\frac{k^5}{k_p^5}\left(1 - \frac{(\varepsilon_L+1)}{p^2} - p^5 - p^3(\varepsilon_L - 1)\right). \end{gathered} \tag{A2}$$

При обозначении $\sin(\varphi) = -b/\sqrt{a^2+b^2}$ имеем формулы интегрирования [33]

$$I_{n(s,c)}(x)=\int x^n \begin{Bmatrix}\sin(bx)\\ \cos(bx)\end{Bmatrix}\exp(-ax)dx=\frac{x^n\exp(-ax)}{a^2+b^2}\left[-a\begin{Bmatrix}\sin(bx)\\ \cos(bx)\end{Bmatrix}\mp b\begin{Bmatrix}\cos(b)\\ \sin(b)\end{Bmatrix}\right]- \\ -\frac{n}{a^2+b^2}\int x^{n-1}\left[-a\begin{Bmatrix}\sin(bx)\\ \cos(bx)\end{Bmatrix}\mp b\begin{Bmatrix}\cos(b)\\ \sin(b)\end{Bmatrix}\right]\exp(-ax)dx, \quad (A2)$$

$$I_{1(s,c)}(x)=\int x\begin{Bmatrix}\sin(bx)\\ \cos(bx)\end{Bmatrix}\exp(-ax)dx=\frac{\exp(-ax)}{a^2+b^2}\left[ax-\frac{a^2-b^2}{a^2+b^2}\begin{Bmatrix}\sin(bx)\\ \cos(bx)\end{Bmatrix}\mp \right. \\ \left. \mp\left(bx+\frac{2ab}{a^2+b^2}\right)\begin{Bmatrix}\cos(bx)\\ \sin(bx)\end{Bmatrix}\right], \quad (A3)$$

$$I_{2(s,c)}(x)=\int x^2\begin{Bmatrix}\sin(bx)\\ \cos(bx)\end{Bmatrix}\exp(-ax)dx=\frac{\exp(-ax)}{a^2+b^2}\left[ax^2-2\frac{a^2-b^2}{a^2+b^2}x-\frac{2a(a^2-3b^2)}{(a^2+b^2)^2}\begin{Bmatrix}\sin(bx)\\ \cos(bx)\end{Bmatrix}\mp \right. \\ \left. \mp\left(bx^2+\frac{4ab}{a^2+b^2}x+\frac{2b(3a^2-b^2)}{(a^2+b^2)^2}\right)\begin{Bmatrix}\cos(bx)\\ \sin(bx)\end{Bmatrix}\right], \quad (A4)$$

$$\int x^n\begin{Bmatrix}\sin(bx)\\ \cos(bx)\end{Bmatrix}\exp(-ax)dx=\frac{\exp(-ax)}{n+1}\sum_{k=1}^{n+1}(-1)^{k-1}\frac{(n+1)n...(n-k+2)}{(a^2+b^2)^{k/2}}x^{n-k+1}\begin{Bmatrix}\sin(bx+k\varphi)\\ \cos(bx+k\varphi)\end{Bmatrix}. \quad (A5)$$

Эти формулы позволяют вычислять интегралы от экспоненты со степенями, если в нижней формуле положить $b$=0. Однако для интегралов такого типа

$$f_n(a,b,c)=\int_b^c x^n\exp(-ax)dx$$

удобны более простые формулы

$$f_{n`}(a,b,c)=-\exp(-ax)\left[\frac{x^n}{a}+\frac{nx^{n-1}}{a^2}+...+\frac{n!}{a^n}+\frac{n!}{a^{n+1}}\right]_{x=b}^{x=c}. \quad (A6)$$

Имеем следующую формулу интегрирования при произвольном $\nu$>0 [33]

$$f_n(a,b,c)=\int_0^\infty x^\nu\exp(-ax)dx=\frac{\Gamma(\nu+1)}{a^{\nu+1}}. \quad (A7)$$

Для четных степеней КО и их производных имеем формулы

$$r_{e,h}^{2m}(k,1)=\left(\frac{1-\varepsilon^{-1/2}(k)}{1+\varepsilon^{-1/2}(k)}\right)^{2m}, \quad (A8)$$

$$\partial_p r_e^{2m}(k,p)=2m\left[\frac{1-\sqrt{\varepsilon(k)+p^2-1}/(p\varepsilon(k))}{1+\sqrt{\varepsilon(k)+p^2-1}/(p\varepsilon(k))}\right]^{2m-1}\partial_p\left[\frac{p\varepsilon(k)-\sqrt{\varepsilon(k)+p^2-1}}{p\varepsilon(k)+\sqrt{\varepsilon(k)+p^2-1}}\right],$$

$$\partial_p\left[\frac{p\varepsilon(k)-\sqrt{\varepsilon(k)+p^2-1}}{p\varepsilon(k)+\sqrt{\varepsilon(k)+p^2-1}}\right]=\frac{2}{\varepsilon(k)}\frac{\sqrt{\varepsilon(k)+p^2-1}-p^2/\sqrt{\varepsilon(k)+p^2-1}}{\left(p+\sqrt{\varepsilon(k)+p^2-1}/\varepsilon(k)\right)^2},$$

$$\partial_p\left[\frac{p\varepsilon(k)-\sqrt{\varepsilon(k)+p^2-1}}{p\varepsilon(k)+\sqrt{\varepsilon(k)+p^2-1}}\right]_{p=1}=2\varepsilon^{-1/2}(k)\frac{1-\varepsilon^{-1/2}(k)}{1+\varepsilon^{-1/2}(k)},$$

$$\partial_p r_e^{2m}(k,1)=\frac{4m}{\varepsilon^{1/2}(k)}\frac{\left(1-\varepsilon^{-1/2}(k)\right)^{2m}}{\left(1+\varepsilon^{-1/2}(k)\right)^{2m}}\,. \tag{A9}$$

$$\partial_p r_h^{2m}(k,p)=2m\left[\frac{1-p/\sqrt{\varepsilon(k)+p^2-1}}{1+p/\sqrt{\varepsilon(k)+p^2-1}}\right]^{2m-1}\partial_p\left[\frac{1-p/\sqrt{\varepsilon(k)+p^2-1}}{1+p/\sqrt{\varepsilon(k)+p^2-1}}\right],$$

$$\partial_p\left[\frac{1-p/\sqrt{\varepsilon(k)+p^2-1}}{1+p/\sqrt{\varepsilon(k)+p^2-1}}\right]=\frac{-2}{\sqrt{\varepsilon(k)+p^2-1}\left(1+p/\sqrt{\varepsilon(k)+p^2-1}\right)^2},$$

$$\partial_p\left[\frac{1-p/\sqrt{\varepsilon(k)+p^2-1}}{1+p/\sqrt{\varepsilon(k)+p^2-1}}\right]_{p=1}=\frac{-2}{\varepsilon^{1/2}(k)\left(1+\varepsilon^{-1/2}(k)\right)^2}, \tag{A10}$$

$$\partial_p r_h^{2m}(k,1)=-\frac{4m}{\varepsilon^{1/2}(k)}\frac{\left(1-\varepsilon^{1/2}(k)\right)^{2m-1}}{\left(1+\varepsilon^{1/2}(k)\right)^{2m+1}}\,. \tag{A11}$$

Для вторых производных получаем

$$\partial_p^2 r_e^{2m}(k,p)=2m(2m-1)\left[\frac{1-\sqrt{\varepsilon(k)+p^2-1}/(p\varepsilon(k))}{1+\sqrt{\varepsilon(k)+p^2-1}/(p\varepsilon(k))}\right]^{2m-2}$$

$$\left\{\partial_p\left[\frac{p\varepsilon(k)-\sqrt{\varepsilon(k)+p^2-1}}{p\varepsilon(k)+\sqrt{\varepsilon(k)+p^2-1}}\right]\right\}^2+2m\left[\frac{1-\sqrt{\varepsilon(k)+p^2-1}/(p\varepsilon(k))}{1+\sqrt{\varepsilon(k)+p^2-1}/(p\varepsilon(k))}\right]^{2m-1}\times,$$

$$\times\partial_p^2\left[\frac{p\varepsilon(k)-\sqrt{\varepsilon(k)+p^2-1}}{p\varepsilon(k)+\sqrt{\varepsilon(k)+p^2-1}}\right]$$

$$\partial_p^2\left[\frac{p\varepsilon(k)-\sqrt{\varepsilon(k)+p^2-1}}{p\varepsilon(k)+\sqrt{\varepsilon(k)+p^2-1}}\right]=\frac{2p}{\varepsilon(k)\sqrt{\varepsilon(k)+p^2-1}\left(p+\sqrt{\varepsilon(k)+p^2-1}/\varepsilon(k)\right)^2}\times$$

$$\times\left[-1+\frac{2p^2}{\left(\varepsilon(k)+p^2-1\right)}-2\frac{\left(\varepsilon(k)-1\right)}{\left(p+\sqrt{\varepsilon(k)+p^2-1}/\varepsilon(k)\right)}\left(1+\frac{p}{\sqrt{\varepsilon(k)+p^2-1}\varepsilon(k)}\right)\right]$$

$$\partial_p^2\left[\frac{p\varepsilon(k)-\sqrt{\varepsilon(k)+p^2-1}}{p\varepsilon(k)+\sqrt{\varepsilon(k)+p^2-1}}\right]_{p=1}=\frac{4-4\varepsilon^2(k)-4\varepsilon^{1/2}(k)+2\varepsilon(k)+4\varepsilon^{-1/2}(k)}{\varepsilon^{5/2}(k)\left(1+\varepsilon^{-1/2}(k)\right)^3}, \tag{A12}$$

$$\begin{gathered}\partial_p^2 r_e^{2m}(k,1)=\frac{8m(2m-1)}{\varepsilon(k)}\left[\frac{1-\varepsilon^{-1/2}(k)}{1+\varepsilon^{-1/2}(k)}\right]^{2m}+\\ +4m\frac{\left(1-\varepsilon^{-1/2}(k)\right)^{2m-1}}{\left(1+\varepsilon^{-1/2}(k)\right)^{2m+2}}\frac{2\varepsilon^{-2}(k)-2-2\varepsilon^{-3/2}(k)+\varepsilon^{-1}(k)+2\varepsilon^{-5/2}(k)}{\varepsilon^{1/2}(k)},\end{gathered} \tag{A13}$$

$$\partial_p^2 r_h^{2m}(k,p) = 2m(2m-1)\left[\frac{1-p/\sqrt{\varepsilon(k)+p^2-1}}{1+p/\sqrt{\varepsilon(k)+p^2-1}}\right]^{2m-2}\left\{\partial_p\left[\frac{1-p/\sqrt{\varepsilon(k)+p^2-1}}{1+p/\sqrt{\varepsilon(k)+p^2-1}}\right]\right\}^2 + \\ +2m\left[\frac{1-p/\sqrt{\varepsilon(k)+p^2-1}}{1+p/\sqrt{\varepsilon(k)+p^2-1}}\right]^{2m-1}\partial_p^2\left[\frac{1-p/\sqrt{\varepsilon(k)+p^2-1}}{1+p/\sqrt{\varepsilon(k)+p^2-1}}\right],$$

$$\partial_p^2\left[\frac{1-p/\sqrt{\varepsilon(k)+p^2-1}}{1+p/\sqrt{\varepsilon(k)+p^2-1}}\right] = \frac{2}{\left(\varepsilon(k)+p^2-1\right)\left(1+p/\sqrt{\varepsilon(k)+p^2-1}\right)^4}\times \\ \times\left[\frac{p\left(1+p/\sqrt{\varepsilon(k)+p^2-1}\right)^2}{\sqrt{\varepsilon(k)+p^2-1}}+\frac{2\left(\sqrt{\varepsilon(k)+p^2-1}+p\right)\left(\varepsilon(k)-1\right)}{\left(\sqrt{\varepsilon(k)+p^2-1}\right)^3}\right],$$

$$\partial_p^2\left[\frac{1-p/\sqrt{\varepsilon(k)+p^2-1}}{1+p/\sqrt{\varepsilon(k)+p^2-1}}\right]_{p=1} = \frac{6\varepsilon^{-1}(k)+4\varepsilon^{-1/2}(k)-2\varepsilon^{-2}}{\varepsilon^{1/2}(k)\left(1+\varepsilon^{-1/2}(k)\right)^4}. \tag{A14}$$

$$\partial_p^2 r_h^{2m}(k,1) = \frac{4m}{\varepsilon(k)}\frac{\left(1-\varepsilon^{-1/2}(k)\right)^{2m-2}}{\left(1+\varepsilon^{-1/2}(k)\right)^{2m+2}}\times \\ \times\left[4m-2+\frac{1-\varepsilon^{-1/2}(k)}{1+\varepsilon^{-1/2}(k)}\left(2+3\varepsilon^{-1/2}(k)-\varepsilon^{-3/2}\right)\right]. \tag{A15}$$

Имеем $r_{e,h}^{2m}(0,1)=1$, $\partial_p r_{e,h}^{2m}(0,1)=0$, $\partial_p^2 r_\alpha^{2m}(0,1)=0$.

Поправочные члены к (33) и (34) имеют вид

$$\delta I_{em}(k,d) = 8m^2\exp(-2mkd)\times \\ \times\left[\frac{\alpha_{e0}^2(k)}{2mkd}+2\alpha_{e0}(k)\alpha_{\alpha1}(k)\left(\frac{1}{2mkd}+\frac{2}{(2mkd)^2}+\frac{2}{(2mkd)^3}\right)+\right. \\ \left.+\alpha_{e1}^2(k)\left(\frac{1}{2mkd}+\frac{4}{(2mkd)^2}+\frac{12}{(2mkd)^3}+\frac{24}{(2mkd)^4}+\frac{24}{(2mkd)^5}\right)\right], \tag{A16}$$

$$\delta I_{hm}(k,d) = 8m^2\exp(-2mkd)\times \\ \times\left[\alpha_{eh}^2(k)\left(\frac{1}{2mkd}+\frac{4}{(2mkd)^2}+\frac{12}{(2mkd)^3}+\frac{24}{(2mkd)^4}+\frac{24}{(2mkd)^5}\right)+\right. \\ +2\alpha_{e0}(k)\alpha_{\alpha1}(k)\left(\frac{1}{2mkd}+\frac{6}{(2mkd)^2}+\frac{30}{(2mkd)^3}+\frac{120}{(2mkd)^4}+\right. \\ \left.+\frac{360}{(2mkd)^5}+\frac{720}{(2mkd)^6}+\frac{720}{(2mkd)^7}\right)+\alpha_{h1}^2(k)\left(\frac{1}{2mkd}+\frac{8}{(2mkd)^2}+\right. \\ +\frac{56}{(2mkd)^3}+\frac{336}{(2mkd)^4}+\frac{1680}{(2mkd)^5}+\frac{6720}{(2mkd)^6}+\frac{20160}{(2mkd)^7}+ \tag{A17}$$

$$+\frac{40320}{(2mkd)^8}+\frac{40320}{(2mkd)^9}\Bigg)\Bigg]\Bigg].$$

Для вычисления поправок имеем

$$\sum_{m=1}^{\infty}\int_0^{\infty}k^3I_{em}(k,d)dk=\left[\varsigma(4)\frac{\Gamma(3)+2\Gamma(2)+2\Gamma(1)}{(2d)^4}-\frac{4}{k_p}\varsigma(4)\left(\frac{\Gamma(4)+\Gamma(3)}{(2d)^5}\right)-\right.$$
$$\left.-\frac{2}{k_p^3}\varsigma(6)\frac{\Gamma(6)+3\Gamma(5)+6\Gamma(4)+6\Gamma(3)}{(2d)^7}\right], \quad \text{(A18)}$$

$$\sum_{m=1}^{\infty}\int_0^{\infty}k^3I_{hm}(k,d)dk=\left\{\varsigma(4)\frac{\Gamma(3)+2\Gamma(2)+2\Gamma(1)}{(2d)^4}-\frac{4\varsigma(4)}{k_p}\left[\frac{\Gamma(4)+3\Gamma(3)+6\Gamma(2)+6\Gamma(1)}{(2d)^5}\right]+\right.$$
$$\left.+\frac{2\varsigma(6)}{k_p^3}\left[\frac{\Gamma(6)+5\Gamma(5)+20\Gamma(4)+60\Gamma(3)+120\Gamma(2)+120\Gamma(1)}{(2d)^7}\right]\right\}. \quad \text{(A19)}$$

$$\int_0^{\infty}k^3\delta I_{em}(k,d)dk=\frac{8\varsigma(4)\Gamma(5)}{m^6(2d)^6k_p^2}+8\varsigma(6)\frac{\Gamma(7)+2\Gamma(6)+2\Gamma(5)}{k_p^4m^8(2d)^8}+$$
$$+8\varsigma(8)\frac{\Gamma(9)+4\Gamma(8)+12\Gamma(7)+24\Gamma(6)+24\Gamma(5)}{4k_p^6m^{10}(2d)^{10}}. \quad \text{(A20)}$$

$$\sum_{n=1}^{\infty}\int_0^{\infty}k^3\delta I_{hm}(k,d)dk=8\varsigma(4)\frac{\Gamma(5)+4\Gamma(4)+12\Gamma(3)+48}{k_p^2m^6(2d)^6}+$$
$$+8\varsigma(6)\frac{\Gamma(7)+6\Gamma(6)+30\Gamma(5)+120\Gamma(4)+360\Gamma(3)+1440}{k_p^4(2d)^8}+ \quad . \quad \text{(A21)}$$
$$8\varsigma(8)\frac{\Gamma(9)+8\Gamma(8)+56\Gamma(7)+336\Gamma(6)+1680\Gamma(5)+6720\Gamma(4)+20160\Gamma(3)+80640}{4k_p^6(2d)^{10}}$$

Коэффициенты в разложении (34) принимают вид

$$a_1=\frac{\Gamma(4)+\Gamma(3)+\Gamma(4)+3\Gamma(3)+6\Gamma(2)+6\Gamma(1)}{\Gamma(3)+2\Gamma(2)+2\Gamma(1)}=\frac{16}{3},$$
$$a_2=\frac{2\Gamma(5)+4\Gamma(4)+12\Gamma(3)+48}{\Gamma(3)+2\Gamma(2)+2\Gamma(1)}=24, \quad \text{(A22)}$$
$$a_3=c_3=\frac{10\varsigma(6)}{\varsigma(4)}(2\varepsilon_L-1)+\frac{1950\varsigma(5)-1200\varsigma(6)}{7\varsigma(4)}.$$

Figures

**On Casimir force between two real metallic plates**

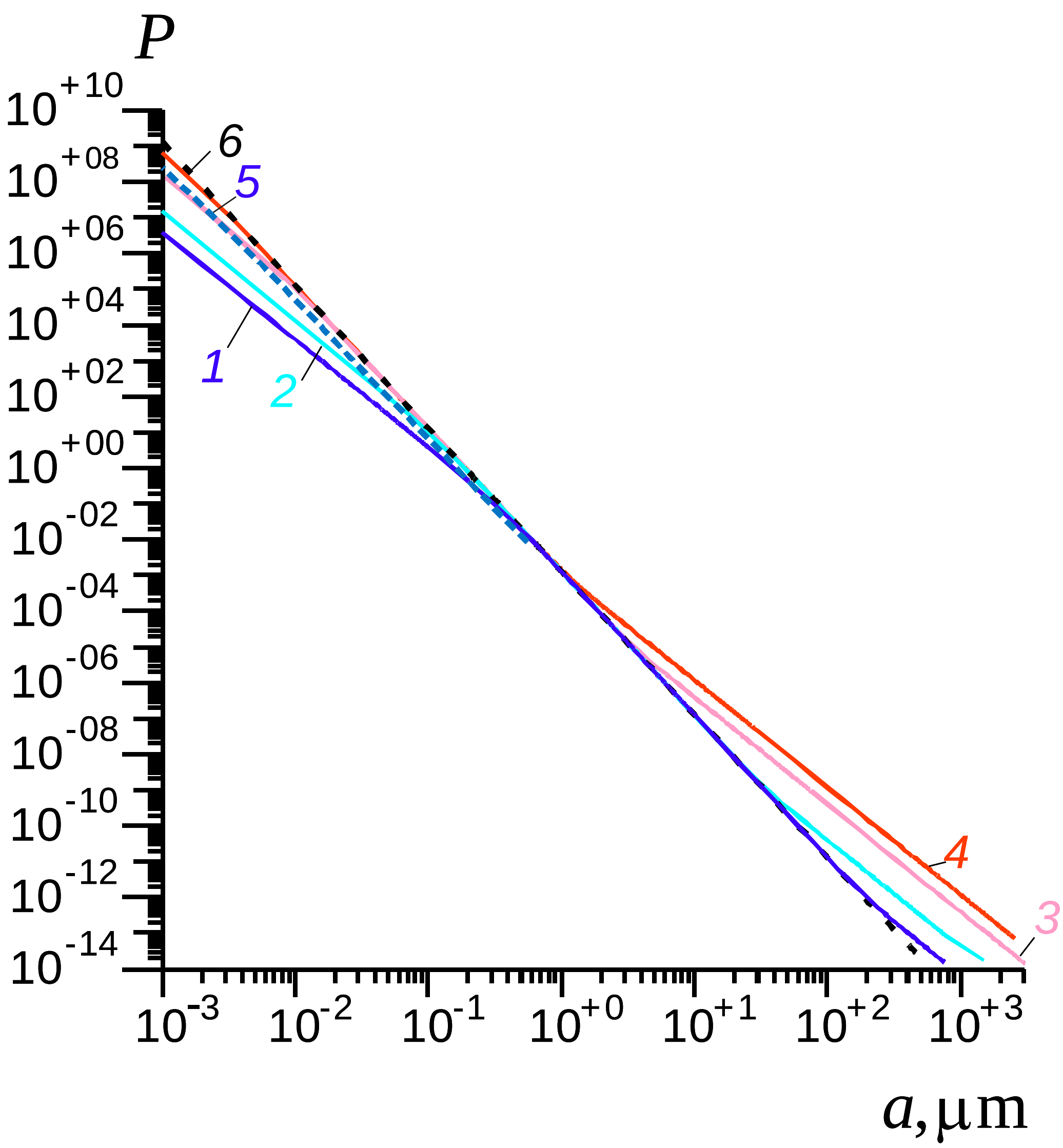

Fig. 1. External Casimir pressure $P$ (N/m$^2$) on two ideally conductive metal layers depending on $a$ (in μm) at different temperatures (formula (6)): 5 K (curve 1), 30 K (2), 300 K (3) and 900 K (4). The curve 5 is the correction of (2) using (44). Casimir's result is a dashed line 6

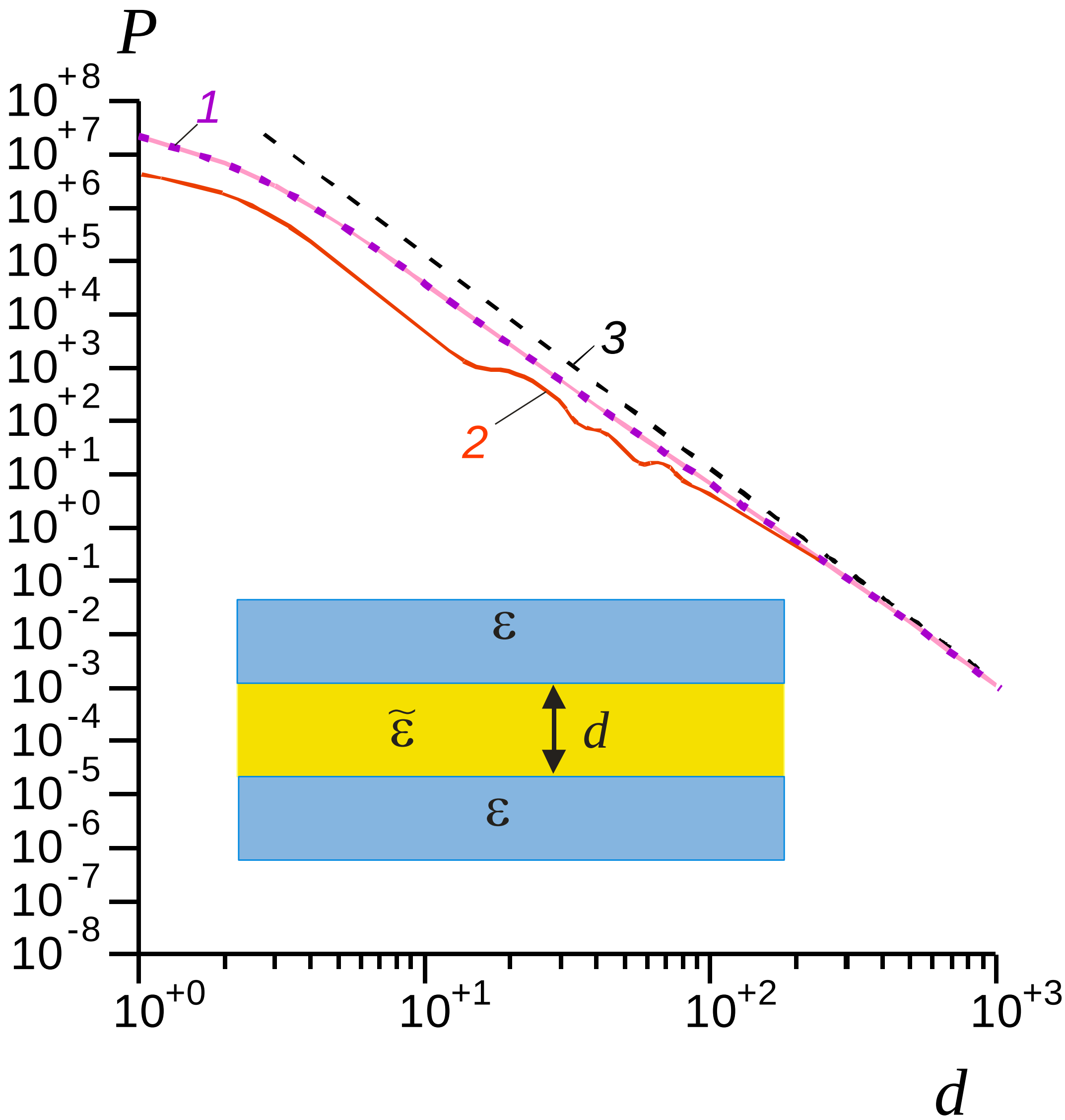

Fig. 2. External Casimir pressure $P$ ($N/m^2$) as a function of thickness $d$ (nm) for thick metal plates (silver, $\omega_p$=1.57×$10^{16}$ Hz, $\omega_c$=3.2×$10^{13}$ Hz, $\varepsilon_L$=9.6) separated by a vacuum gap $d$ (curve 1) and approximation $\varepsilon_L$ by three Lorentz terms with frequencies with frequencies of 2×$10^{17}$, 5×$10^{17}$ and $10^{18}$ Hz (2). The result of Casimir is shown by line 3

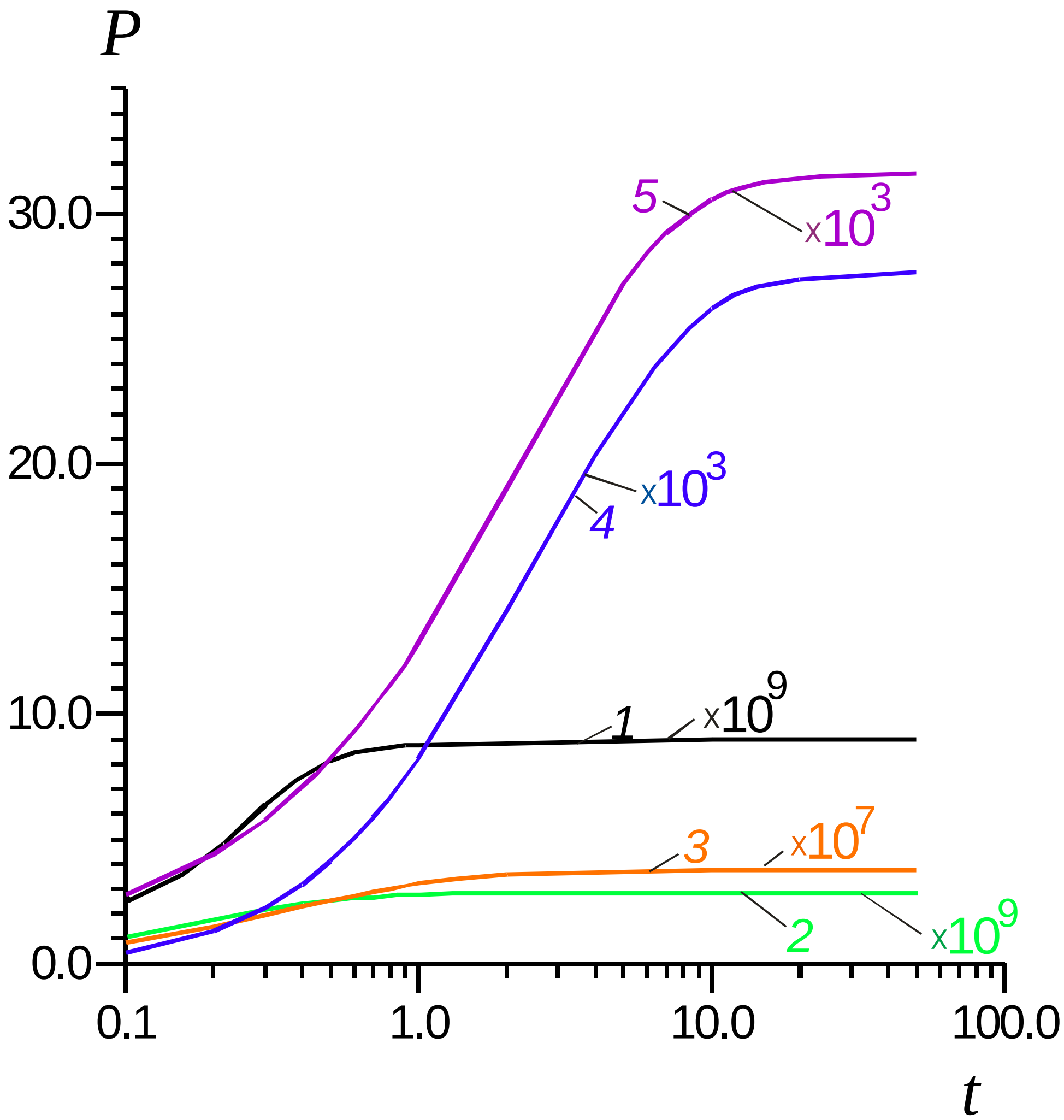

Fig. 3. External Casimir pressure $P=F/L^2$ (N/m$^2$) on two dielectric layers $\varepsilon_L$=20 depending on the thickness of the layers t (nm) at different distances *d* (nm): *d*=0.01 (curve 1); 0.1 (2); 1 (3); 10 (4,5). Curves 1–4 are plotted in the absence of conductivity $\omega_p = 0$, curve 5 – in its presence for DM ($\omega_p$=1.5×10$^{16}$ Hz, $\omega_c$=10$^{13}$ Hz)

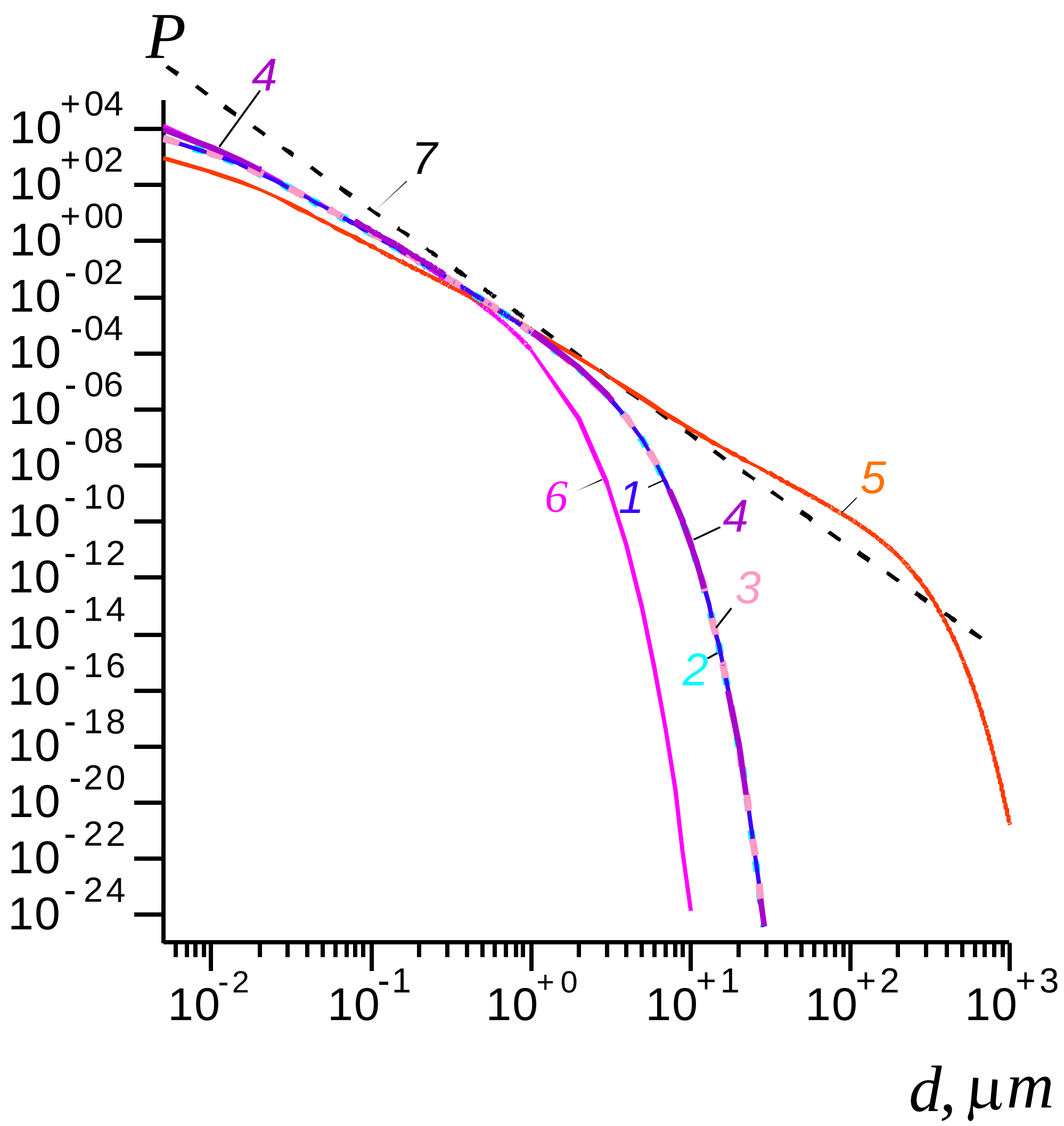

Fig. 4. External Casimir pressure $P=F/L^2$ ($N/m^2$) on two thick metal layers depending on the distance $d$ for different DP models: plasma (curves 1.5), DM (2), DSM (3) Onsager-Drude-Smith (4,6). Curves 1–4 are plotted for $T$=300 K, curve 5 for $T$=5 K, curve 6 for $T$=900 K. Dashed line 7 is the result of Casimir ($T$=0)